\documentclass[journal]{IEEEtran}

\usepackage{color}
\usepackage{cite}
\usepackage[linesnumbered,ruled,vlined]{algorithm2e} 
\usepackage{amsmath}
\usepackage{amssymb}
\usepackage{amsthm}
\usepackage{amscd}
\usepackage{graphicx}%
\usepackage[font=small,skip=2pt]{caption}
\usepackage{array}
\usepackage{float}

\usepackage[switch,pagewise]{lineno}

\usepackage[hyphens]{url}

\usepackage[table]{xcolor}
\definecolor{algCommColor}{rgb}{0.0, 0.0, 0.61}

\begin{document}

\title{The Fog Development Kit: A Platform for the Development and Management of Fog Systems} 

\twocolumn

\author{\IEEEauthorblockN{Colton Powell,
        Christopher Desiniotis,
        and Behnam Dezfouli}\\
        \IEEEauthorblockA{Internet of Things Research Lab, Department of Computer Science \& Engineering, Santa Clara University, USA}\\
        {\small\texttt{ctpowell@scu.edu, cdesiniotis@scu.edu, bdezfouli@scu.edu}}
}

\markboth{SCU Internet of Things Research Lab (SIOTLAB)}%
{Powell \MakeLowercase{\textit{et al.}}: The Fog Development Kit}

\maketitle

\pagenumbering{arabic}

\begin{abstract}
With the rise of the Internet of Things (IoT), fog computing has emerged to help traditional cloud computing in meeting scalability demands. 
Fog computing makes it possible to fulfill real-time requirements of applications by bringing more processing, storage, and control power geographically closer to end-devices.
However, since fog computing is a relatively new field, there is no standard platform for research and development in a realistic environment, and this dramatically inhibits innovation and development of fog-based applications.
In response to these challenges, we propose the \textit{Fog Development Kit} (FDK).
By providing high-level interfaces for allocating computing and networking resources, the FDK abstracts the complexities of fog computing from developers and enables the rapid development of fog systems.
In addition to supporting application development on a physical deployment, the FDK supports the use of emulation tools (e.g., GNS3 and Mininet) to create realistic environments, allowing fog application prototypes to be built with zero additional costs and enabling seamless portability to a physical infrastructure.
Using a physical testbed and various kinds of applications running on it, we verify the operation and study the performance of the FDK.
Specifically, we demonstrate that resource allocations are appropriately enforced and guaranteed, even amidst extreme network congestion.
We also present simulation-based scalability analysis of the FDK versus the number of switches, the number of end-devices, and the number of fog-devices.
\end{abstract}



\IEEEpeerreviewmaketitle

\section{Introduction}
In today's world of smart cars, smart cities, smart homes, Industry 4.0, and mobile healthcare, almost every device is connected to the Internet. 
Whether they be televisions, sensors, or wearable devices, these technologies often generate data and require computation and storage needs that cannot be met at the network edge.
With the growing number of interconnected devices and IoT applications arises the challenge of handling a massive amount of data in a highly efficient manner.

Cloud computing offers a partial solution to this dilemma by providing massive infrastructure and powerful applications.
However, cloud computing is not suitable for real-time and mission-critical application domains with stringent runtime and latency requirements.
Additionally, cloud computing cannot scale sufficiently to handle the processing, storage, and communication demands of billions of IoT devices \cite{fog2016survey,Bonomi2012,shi2016promise,cao2015distributed}. 
Fog computing aims to solve this challenge by bringing additional computing, storage, and control capabilities to the network edge. 
The increased number of powerful computation and networking platforms has made the implementation of fog architectures a worthwhile undertaking \cite{Amirtharaj2019}. 
Fog is intended to work alongside the cloud, forming a things-fog-cloud continuum where applications can be served promptly \cite{fog2016survey}.

By using the things-fog-cloud continuum, requests generated by end-devices (things) can be serviced in the fog, thereby avoiding transmission to the cloud and significantly reducing packet latency and network congestion. 
Resource-constrained devices such as medical devices can offload computation- and communication-intensive tasks to nearby fog-devices to meet real-time constraints. 
For example, consider a scenario where medical devices in a hospital monitor patients.
Once a device detects an anomaly, it can request resources from fog-devices for further processing and real-time results.
We refer to these systems as \textit{fog systems}, where applications on end-devices may offload their computational tasks to nearby fog-devices.
These systems may optionally connect to cloud data centers for increased accuracy in the decision-making process.

There exist significant obstacles for research and development in the realm of fog systems.
\textit{First}, end-devices need to request and \textit{reserve} resources to meet the Quality of Service (QoS) demands of underlying applications, meaning that any efficient fog system must operate with a resource allocator. 
Traditional load balancers are not sufficient in fulfilling the needs of heterogeneous IoT applications, where end-devices require \textit{guaranteed} resources to meet their stringent runtime and latency requirements. 
While many resource allocation platforms have been proposed \cite{enorm, yin2018, Skarlat2016resourceprovisioning}, few systems allocate both networking and computing resources.
Furthermore, to the best of our knowledge, no such platforms have integrated software-defined networking (SDN) into their architecture, where fog-device resource allocation, network bandwidth allocation, and customizable routing policies are all consolidated into a single, comprehensive platform.
\textit{Second}, most of the existing works employ simulation to evaluate the efficiency of their resource management approaches \cite{Skarlat2016resourceprovisioning, ansari2018, yin2018, cloudsimsdn,containercloudsim,ifogsim}; thereby highlighting an apparent lack of development tools for research in this field.
In order to exhaustively test new approaches in realistic environments, and to accelerate research in fog computing, a standard research and development platform is needed.
\textit{Finally}, it can be quite expensive to prototype and test the performance of a real fog-based application.
For example, creating even the most straightforward application requires constructing an infrastructure of end-devices, fog-devices, and networking hardware, which can be costly.
Therefore, the creation of \textit{complex} software components and a \textit{costly} physical infrastructure must precede the development of such applications.
This combination of complexity and cost poses an immense barrier of entry for researchers and engineers.
Since fog computing is still in its infancy, there is no standard development kit or platform which has solved all of these issues in the form of a single, complete development package.
Without such a platform, the advancement of pertinent, real-time applications will be slow, given the barriers of entry.

In this paper, we set out to address this problem by proposing the \textit{Fog Development Kit} (FDK)\footnote{The FDK is accessible at the following address: https://github.com/SIOTLAB/Fog-Development-Kit.git}: A development and management platform for fog systems.
The FDK is intended to bring together all of the elements of fog computing into one comprehensive framework, where developers can begin building fog-based applications with ease and without all of the barriers mentioned above.

The FDK addresses development complexity by providing a cutting-edge resource allocation scheme, which supports any arbitrary fog-based application running on top of it.
Specifically, by integrating SDN and virtualization technologies, the FDK enables end-devices to utilize its messaging protocol to request for computing and communication resources.
If sufficient resources are available, the FDK instantiates a container in a fog-device with the desired computing resources, finds an efficient path through the network for communication between the end-device and the fog-device, and allocates the requested bandwidth along the identified path.
The complexity of resource allocation is thus handled by the FDK.
For example, suppose a developer plans to build a facial recognition system, where resource-constrained end-devices connected to cameras live-stream video data to fog-devices for heavy-duty processing.
Here, it is only required to develop an application for the end-device to collect and stream video data, as well as the containerized services running on fog-devices to receive and process the data.
The FDK handles all of the underlying system complexities such as managing computational resources of fog-devices, path reservation, and bandwidth slicing between end-devices and fog-devices.

The FDK supports developing applications in both physical and emulated environments.
Built on top of OpenDaylight (ODL) \cite{odl}, the FDK utilizes standard SDN protocols to communicate with physical network devices.
Moreover, the FDK is designed to be used in unison with Open vSwitch (OVS) \cite{ovs}, which performs network resource allocation using the OVSDB management protocol \cite{ovsdb} and enforces data flow routing using the OpenFlow protocol \cite{openflow13}.
Therefore, in addition to supporting physical environments, the FDK was designed to be used with emulation technologies so that developers could leverage tools such as GNS3 \cite{GNS3} and Mininet \cite{mininet} to prototype fog-based applications.
GNS3 and Mininet provide the capability of emulating network topologies on a personal computer.
These tools allow virtual machines and containers running on the computer to communicate with each other in a virtualized environment.
With this, the FDK can run on a completely emulated network consisting of Linux virtual machines (VM) serving as end-devices and fog-devices, and OVS VMs which handle the messages exchanged between these devices.
Therefore, the FDK enables the development of applications in an emulated environment at zero additional cost.
In addition, any applications developed on top of the FDK can be ported from an emulated environment to a physical infrastructure.


We evaluate the correctness and performance of the FDK by using a physical testbed consisting of eight end-devices, four fog-devices, and five OpenFlow switches.
Our results show that resource allocation and deallocation delays are less than 279 ms and 256 ms, respectively, for 95\% of transactions.
We also show the resiliency of the FDK by analyzing the impact of various network conditions and levels of congestion on already-running application transmission speeds and show that bandwidth allocations are accurately enforced and upheld regardless of network conditions.
In addition, we present a simulation-based scalability analysis to demonstrate the impact of network size, topology type, number of end-devices, and number of fog-devices on controller overhead and communication delay.


The rest of this paper is organized as follows.
We present the related work in Section \ref{RelatedWork}.
In Section \ref{FogDevelopmentKit}, we summarize the goals and features of the FDK.
Section \ref{SystemArchitecture} presents the system architecture and operation of the FDK.
In Section \ref{Evaluation}, we present performance evaluation using a physical testbed and simulation.
In Section \ref{FutureWork}, we highlight potential future work, and lastly in Section \ref{Conclusion} we conclude the paper.

\section{Related Work} 
\label{RelatedWork}
In this section, we overview relevant simulation platforms and justify the importance of the FDK.
Also, we summarize existing load balancing and resource allocation schemes and identify their shortcomings when applied to fog-based applications.
Finally, we investigate other existing fog architectures and platforms, and highlight the benefits that the FDK holds over these alternatives.

\subsection{Simulation Platforms}


Due to the significant cost of creating fog and cloud network infrastructures, simulation-based study is the most widely-used approach to evaluate the performance of proposed mechanisms \cite{Skarlat2016resourceprovisioning, ansari2018, yin2018}.

CloudSim \cite{cloudsim} is perhaps the most popular cloud simulation platform available, which is used for modeling the cloud and application provisioning environments. 
It is a discrete event-based simulator written in Java, meaning that it does not actually emulate (virtualize) network entities such as routers and switches.
Instead, CloudSim uses a latency matrix, which contains predefined values for the latency between entities.
Additionally, CloudSim can model dynamic user workloads by exposing a set of methods and variables to configure the resources of simulated VMs.

There are also many extensions to CloudSim, such as CloudSimSDN \cite{cloudsimsdn}, ContainerCloudSim \cite{containercloudsim}, and iFogSim \cite{ifogsim}, which attempt to broaden CloudSim's model to include SDN, container migration simulation, and fog computing, respectively.
However, because CloudSim and these associated extensions are strictly-simulation based, they ultimately do not solve the problems of cost and complexity associated with developing an actual fog application.
Rather, they simply avoid the problem altogether by simulating the entire system. 
Therefore, while CloudSim is a worthy platform for evaluating cloud architectures, load balancing algorithms, etc., it fails to actually serve as a valid fog-based application development platform because projects developed in CloudSim are not portable to a real environment. 
Likewise, the same can be said for most other simulation platforms for similar reasons.
In contrast, the FDK can be used to develop actual fog applications in both physical and emulated environments.
Furthermore, after a fog application is developed in an emulated environment, that application can then be seamlessly ported to a physical environment (and vice versa).

\subsection{Resource Management and Allocation}

Resource management is key to the success of any fog system and consists of two main components: \textit{networking resource management}, and \textit{computational resource management}.

Typically, networking resource management is accomplished using a load balancer, which attempts to find a suitable path to one or more destinations while spreading traffic throughout the network to avoid congestion.
In many cases, Equal-Cost Multi-Path (ECMP) routing is used to manage network resources by distributing traffic throughout the network. 
However, ECMP is congestion oblivious and studies \cite{clove,hermes} suggest that ECMP's performance is far from optimal and that it is known to result in unevenly distributed network flows and poor performance.
In response, Katta \textit{et al.} proposed Clove \cite{clove}, a congestion-aware load balancer that works alongside ECMP by modifying encapsulation packet header fields to manipulate flow paths, ultimately providing lower Flow Completion Times (FCT) than ECMP.
Clove identifies disjoint paths and changes the 5-tuple of the overlay network to distribute traffic over these paths. 
It also uses ECN to detect congestion.
Similarly, Zhang \textit{et al.} proposed Hermes \cite{hermes}, a distributed load balancing system, which offers up to 10\%-20\% faster FCT than Clove. 
While Clove can handle link failures and topology asymmetry, Hermes can handle more advanced and complex uncertainties such as packet black-holes and switch failures.

Unfortunately, load balancers do not adequately fulfill the network resource management requirements of fog systems. 
Load balancers simply find multiple paths for traffic distribution, whereas fog systems need to actually \textit{reserve} bandwidth along paths to fulfill application demands such as real-time exchange of medical monitoring data.

%
%
%
%
%
There are mechanisms that utilize actual network resource allocation to provide timely and reliable services. 
Akella \textit{et al.} \cite{qosNetworkResourceAllocation} proposed a method for guaranteeing network resources and reliable QoS.
They leverage OVS, OVSDB, and SDN technologies to create three tiers of cloud QoS levels, where each tier allocates a specific amount of bandwidth to a user-cloud service.
This is performed by dynamically creating packet queues on switches along the communication path, followed by then creating OpenFlow flows on those switches that enqueue traffic belonging to one of the QoS levels onto the appropriate packet queue.
Kumar \textit{et al.} \cite{endtoendDelayGuarantee} proposed a mechanism to extend SDN infrastructure to be ``delay aware" by finding paths for data flows to ensure end-to-end delays are guaranteed.
To this end, they use a similar scheme where packet queues are dynamically created along a path. 
Then, one flow entry is created and assigned per queue, and all packets belonging to a critical network flow are forwarded to the packet queue associated with that flow.
They also propose a path selection algorithm to meet the desired delay and bandwidth constraints of each flow.


On the other hand, computational resource management often involves the use of VMs and containers, which can be configured to use a specific, limited amount of resources.
The amount of resources allocated to a VM or container directly affects the execution time of tasks and services. 
Therefore, the allocation of these resources is critical in ensuring the timely processing of essential data.
Containers hold an advantage over VMs in the context of resource allocation in the fog, as they tend to be more lightweight and, more importantly, provide finer granularity in allocating resources.
For example, when allocating processing power to VMs, the available options only allow for the specification of the number of entire CPU cores that a particular VM can use.
On the other hand, container technologies like Docker \cite{docker} provide interfaces for specifying more in-depth options when running a container.
For example, container options allow the specification of a fractional number of cores that can be used (e.g., 1.25 CPU cores), in addition to the proportion of CPU cycles that can be utilized, which enables more precise, granular control of resource allocation.

Container management is typically performed through the use of orchestration software, such as Docker Swarm mode \cite{dockerSwarm} or Kubernetes \cite{kubernetes}, which provides functionality for remotely managing, instantiating, and shutting down containers.
These container orchestrators currently serve as the backbone for computing resource allocation in fog and cloud systems, and current research involves more advanced use cases, such as investigating and optimizing live container migration techniques \cite{containerLiveMigration}.
This is critical to the success of such systems, as live migration may interrupt running services, degrading performance and increasing completion delays.
Ansari \textit{et al.} \cite{ansari2018} investigated approaches to resource management and VM migration for fog-based IoT applications in mobile networks. 
They proposed Latency Aware proxy VM Migration (LAM), which solely considers latency when assigning a fog colony to a mobile IoT device, and Energy Aware proxy VM Migration (EAM), which considers the energy consumption of colonies.
They simulated LAM, EAM, and static VM allocation, compared all the three approaches and discussed the tradeoffs involved. 
For simulation, they used EveryWare Lab's user movement trace \cite{everywarelab} to emulate movement patterns of mobile devices. 
However, the authors acknowledge the need to conduct further experiments on physical infrastructure.

\subsection{Fog Architectures and Platforms}
Many fog architectures involving automated resource management have been proposed.
Skarlat \textit{et al.} \cite{Skarlat2016resourceprovisioning} created a resource provisioning system using a fog-cloud middleware component.
The middleware oversees the activity of fog colonies, which are micro data centers consisting of fog cells where tasks and data can be distributed and shared among the cells.
This system merely manages fog computing resources and does not \textit{allocate} those resources, nor does it perform any allocation of network resources.

Yin \textit{et al.} \cite{yin2018} built a novel task-scheduling algorithm and designed a resource reallocation algorithm for fog systems, specifically for real-time, smart manufacturing applications.
However, unlike the previous work, a management software component is not used in their approach, and each fog-device is burdened with the task of deciding whether to accept, reject, or send requests to the cloud.
Resource reallocation is periodically run on a single fog-device, reallocating resources among tasks in order to meet delay constraints.
Their results show reduced task delays and improved resource utilization of fog-devices.
However, their experiments are strictly simulation-based, and the resource management scheme only includes a single fog-device during decision making.


Finally, Wang \textit{et al.} \cite{enorm} proposed a novel resource management framework for edge nodes called ENORM.
Upon startup of the system, an edge manager software installed on all edge nodes gathers and stores available system resources. 
Then, each edge node listens for resource requests from a cloud manager software installed on a cloud server.
Each resource request starts with a handshaking process that eventually leads to the initialization of a fog application.
In contrast, fog-devices in our proposed scheme are managed by a central controller running the FDK.
The FDK then receives service requests from end-devices requesting the instantiation of a fog application service with a specific amount of resources.
If sufficient resources exist, the FDK leverages SDN and containerization technologies to remotely perform both computational and network resource allocation. 

\section{The Fog Development Kit}
\label{FogDevelopmentKit}
The FDK addresses all of the problems mentioned in Section \ref{RelatedWork} by enabling the creation and deployment of fog-based applications in physical and emulated environments.
The FDK also provides a comprehensive SDN-based resource allocation scheme.
This section presents the main features offered by the FDK.

\subsection{Resource Allocation}
The FDK provides comprehensive resource allocation capabilities to ensure that requests made by end-devices are fulfilled completely and in a timely manner.
This is accomplished by providing a resource allocation scheme where both network resources and fog-devices' computational resources can be sliced and allocated.
This automated resource allocation offers several benefits.
First, it ensures that the services requested by end-devices own a dedicated slice of the network, and provides the possibility of guaranteeing network latency and bandwidth for communication with fog-devices.
Second, to guarantee application processing deadlines, it ensures that fog-devices are not overwhelmed by end-devices' requests.
These two features are essential in many fog systems as they ensure expedited processing and seamless interactions between end-devices and fog-devices, which are key advantages that fog computing holds over cloud computing.

\subsection{Agility}
\label{agility}
Working with the FDK does not necessarily require access to a physical testbed.
To support emulated topologies and in order to reduce the development and prototyping costs of fog-based applications, developers can use tools such as GNS3 \cite{GNS3} and Mininet \cite{mininet} to build a complete network of end-devices, fog-devices, and OVS nodes using VMs and containers.
Another VM can be used as the controller, running the FDK and SDN controller software.
The controller VM fulfills the requests made by end-devices by allocating resources and instantiating containerized services in fog-devices.
Therefore, the FDK offers agility to developers by making it possible to quickly begin creating fog-based applications using only a personal computer, while also making the process significantly cheaper.





\subsection{Portability}
Fog-based applications running on emulated topologies may need to be ported over to physical, production topologies once they are complete.
To meet this need, the FDK is designed to be highly portable.
Fog-based applications written on top of the FDK are intended to be portable in their entirety to physical systems.
To satisfy this, the FDK can be installed installed on a virtual or physical Linux machine (acting as a central controller) with Python 3, Docker, ODL, and the necessary ODL plug-ins installed.
In addition, in order to take advantage of the network resource allocation capabilities of the FDK, the switching devices throughout the topology must support the OpenFlow 1.3 and OVSDB protocols.
Considering the wide-spread acceptance of OVS in large-scale environments \cite{fang2018evaluating}, we used OVS as our switch software.
OVS can also be installed on any virtual or physical Linux machine. 
Finally, large vendors such as Cisco and Juniper Networks also carry OpenFlow 1.3 and OVSDB compatible switches \cite{cisco_ovsdb, juniper_ovsdb}, which could allow for a port of the FDK and any fog-based applications developed on top of it to a production-grade physical network.



\subsection{Application Independence}
The key principle that the FDK is designed to fulfill is \textit{application-independence}. 
That is, the FDK aims to support any general fog-based applications being run on top of it, in order to ensure that a variety of heterogeneous services can be developed.
To this end, the FDK provides a \textit{messaging protocol} for end-devices to request resources and instantiate specific containerized applications in the fog-devices to handle their processing needs. 
Conversely, the messaging protocol also provides methods to deallocate resources and terminate containerized applications. 
Therefore, as long as all resource requests follow this protocol, any fog-based application can request resources and leverage the power of fog-devices.
In Section \ref{ResourceManager} we show how various types of applications can be adapted to work with the FDK.

\section{System Architecture}
\label{SystemArchitecture}

\begin{figure}[t]
\centering
\includegraphics[width=1\linewidth]{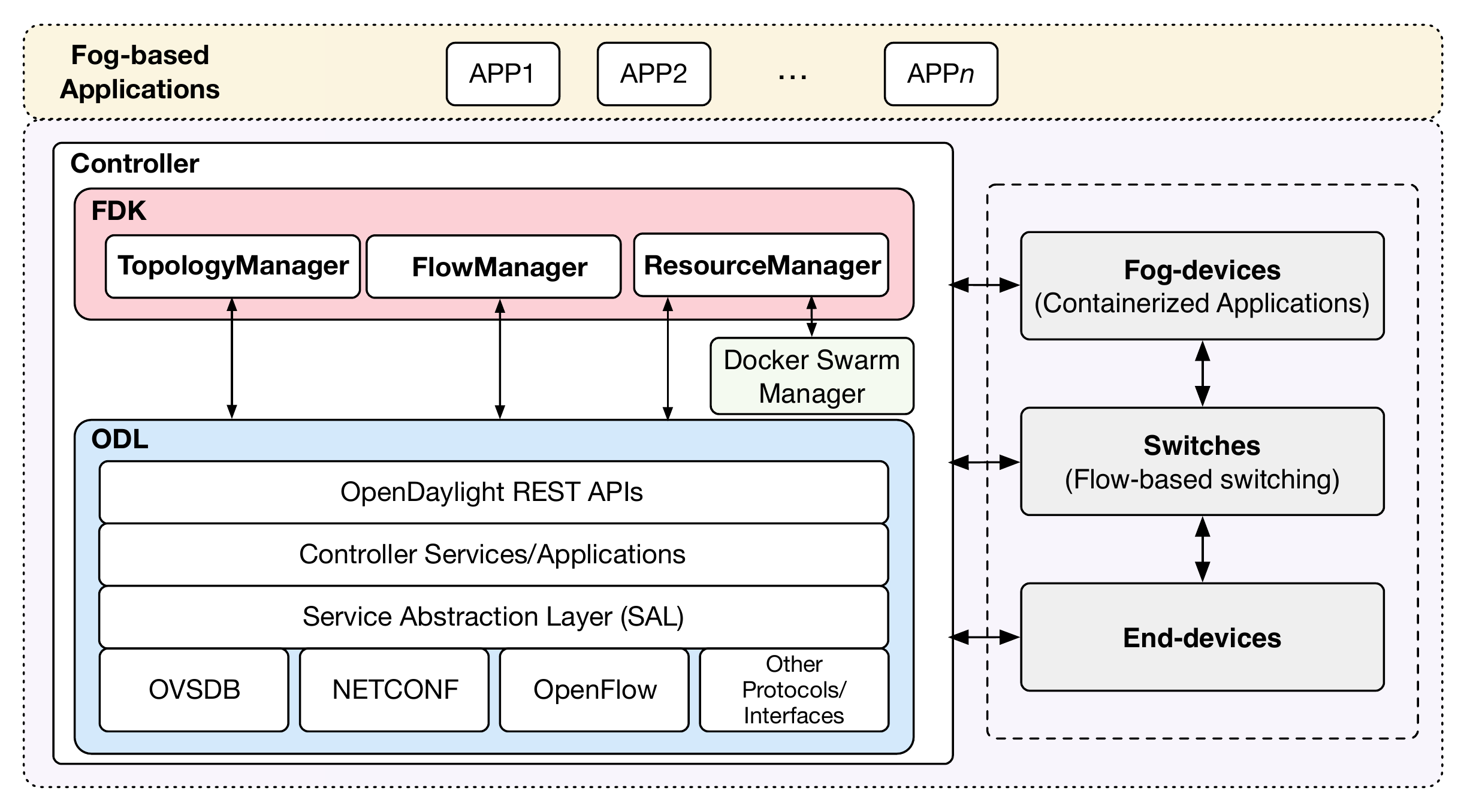}
  \captionof{figure}{Overall system architecture.}
  \label{controller}
\end{figure}

Figure \ref{controller} shows the overall fog system architecture including four major components: \textit{controller}, \textit{end-devices}, \textit{switches}, and \textit{fog-devices}.
End-devices are resource-constrained and cannot completely satisfy application requirements \cite{shi2016promise,cao2015distributed,yi2015survey,fog2016survey}.
Therefore, these devices communicate with more powerful machines---fog-devices---to offload their computing\footnote{Here, \textit{computing} is a general term that includes processing, storage, and communication.} requirements.
For example, an end-device may represent a Raspberry Pi board that captures images and streams them to fog-devices running object recognition algorithms.
Another example of an end-device is a smartphone that collects sensory data from multiple medical devices and transmits them to fog-devices for anomaly detection.
End-devices and fog-devices are connected through switches that support flow-based forwarding.
End-devices, switches, and fog-devices are referred to as \textit{nodes}.
Nodes are monitored and configured by the controller running the FDK.

The FDK itself is a user-space application that operates within the controller and oversees the operation of end-devices, fog-devices, and switches.
The FDK interacts with ODL to control communication paths and manage network resource allocation, and also leverages the Docker containerization technology to remotely instantiate services on fog-devices with a specific amount of resources. 


ODL is an SDN controller software that enables remote management and configuration of networks.
In the case of the FDK, these capabilities are leveraged using ODL's northbound REST interfaces and Model-Driven Service Abstraction Layer (MD-SAL).
At a high level, the MD-SAL allows developers to define data models for ODL software plug-ins and extend the functionality of ODL.
These plug-ins provide additional \textit{northbound} REST APIs.
Invocations of these APIs may utilize a variety of \textit{southbound} network management protocols such as OpenFlow, NETCONF \cite{netconf}, and OVSDB to ultimately configure or modify devices on the network.
These invocations must also include a data body that is in accordance with the YANG data model \cite{yang} defined by the corresponding ODL plugin being utilized.
Upon validation of the data body, it is pushed to the MD-SAL's \textit{configurational data store}, which reflects the desired configuration of the network.
Then, the corresponding plug-in uses the information placed in the configurational data store to apply the desired changes to the appropriate network devices using southbound protocols and interfaces.
Once applied, these changes are reflected in the MD-SAL's \textit{operational data store}, which represents the actual, physical state of the network.
In effect, the MD-SAL supports the development of extensions to ODL, making it an extensible, modular, and versatile SDN controller that has the ability to grow and evolve over time.
In particular, the FDK utilizes ODL's comprehensive set of northbound REST APIs to perform network management using a variety of southbound protocols.
For example, the FDK pushes OpenFlow flows to switches and then remotely configures the switches via OpenFlow 1.3 and OVSDB, respectively, even though the only interfaces accessed by the FDK are ODL's northbound REST APIs.




Docker \cite{docker} is a platform that allows for building, sharing, and executing applications within containers.
Each container is defined by an \textit{image} file, which specifies its exact contents. 
Image files are typically stored in centralized repositories and are accessible by remote compute nodes.
Docker deploys containers by downloading the image file from the remote repository (unless the image is already cached locally) and then instantiates the container using this file.
Docker Swarm mode is a feature that allows for the management and orchestration of such containers on remote machines.
Because these containers have specifiable resource allocation parameters, the FDK leverages Docker Swarm mode to provide fog-device resource allocation capabilities and to instantiate containerized services for end-devices.


The FDK combines and builds upon the functionality of Docker and ODL using \textit{three manager objects} that oversee the entire network and provide interfaces for querying data and manipulating the topology.
These objects are detailed in the rest of this section.

\subsection{TopologyManager}
\label{TopologyManager}

The FDK uses a \textit{TopologyManager} component to query, update, and manage the network topology. 
The core APIs for this component are described in Table \ref{tab:topologyAPIs}.
\begin{table}[t]
    \centering
    \caption{TopologyManager APIs}
    \begin{tabular}{|l|p{4.3cm}|}
    \hline
    \textbf{API} & \textbf{Description} \\ \hline \hline
    \texttt{\scriptsize{update\_topology()}} & Query topology information from ODL and update the topology\\ \hline
    \texttt{\scriptsize{create\_queue()}} & Create/update rate-limited queue on switch\\ \hline
    \texttt{\scriptsize{delete\_queue()}} & Delete queue from switch\\ \hline
    \texttt{\scriptsize{create\_qos()}} & Create QoS entry on switch\\ \hline
    \texttt{\scriptsize{delete\_qos()}} & Delete QoS entry from switch\\ \hline
    \texttt{\scriptsize{place\_queue\_on\_qos()}} & Place queue on QoS entry \\ \hline
    \texttt{\scriptsize{remove\_queue\_from\_qos()}} & Remove queue from QoS entry\\ \hline
    \texttt{\scriptsize{place\_qos\_on\_port()}} & Place QoS entry (containing queues) on switch port \\ \hline
    \texttt{\scriptsize{remove\_qos\_from\_port()}} & Remove QoS entry from switch port \\ \hline
    \end{tabular}
    \label{tab:topologyAPIs}
\end{table}
On startup, the TopologyManager first issues queries to the MD-SAL's operational data store for data pertaining to the ODL OpenFlow plugin, the ODL node inventory, and the OVSDB plugin to gather data on the entire topology.
The results returned by the OpenFlow plugin include information regarding all network devices (i.e., end-devices, fog-devices, switches) and connecting links.
The results returned by the ODL node inventory contain more in-depth information on the OpenFlow switches and their network interfaces, and provide information on the speed of the interfaces and how much data has been transmitted across them since ODL started.
Finally, the results returned by the OVSDB plugin contain information about the configuration of OpenFlow switches as well as the information required to configure them remotely.

The TopologyManager consolidates all the information returned by these calls within a single Topology object, which models the network topology as a graph. 
Links are modeled as directed graph edges, with each one containing multiple data fields such as the current utilization of the link, the current bandwidth allocations on the link, and port identifiers at the endpoints.
The nodes across the network are modeled as end-devices, fog-devices, and switches using a set of device type classes provided within the Topology object. 
The data stored for each node varies depending on its type.
For example, each fog-device object contains information such as the total amount of processing and memory resources on the device, which is later used by the FDK to slice the resources and prevent over-allocation.
Similarly, the OpenFlow switch objects store information regarding their current configurations and the flows installed in their flow tables, which is later used by the FDK to shape network traffic paths and to manage the allocation of communication resources.
Therefore, the TopologyManager serves as a comprehensive directory of information pertaining to the state and structure of the network and the availability of resources across it.

After building the Topology object, the TopologyManager creates a background thread to continuously update the network topology over time.
This thread issues the previously mentioned queries to the ODL operational data store to gather information on the latest state of the topology.
Then, the thread analyzes the differences between the returned data and the current Topology object, and then updates the Topology object to reflect the more recent topology information returned by ODL by making the appropriate changes (such as adding links and/or nodes).

The TopologyManager also provides a large number of APIs for managing OpenFlow switches via the OVSDB management protocol. 
These interfaces provide capabilities for creating and deleting constructs such as packet queues, QoS entries, and ports, which are used by the ResourceManager component of the FDK when allocating network resources.
It should be noted that all OpenFlow data and OVSDB data are originally returned as separate topologies by ODL, and there is no immediately-apparent way to relate data between the two. 
In the case of the OVSDB data, the MAC address of the bridge being controlled by ODL is returned in the query to the OVSDB plugin, which can then be converted to an OpenFlow node ID by stripping out the colons in the MAC address, converting the remaining hex value to a decimal value, and prepending ``\texttt{openflow:}" to the remaining decimal value. 
The FDK then uses this relationship when storing data in Topology objects, and effectively merges the two separate OpenFlow and OVSDB data sets into the single aforementioned Topology object.

Finally, the TopologyManager provides a \textit{greeting server} thread used to handle greeting messages sent by end-devices and fog-devices.
End-devices and fog-devices are configured to send greeting messages upon boot up.
Each message contains a device type and a node ID field, in addition to some supplementary information.
The device type field specifies whether the device is a fog-device or an end-device, and the node ID correlates the device with one that was found in the MD-SAL operational data store.
By building this association via greeting messages, the TopologyManager can identify all of the nodes in the Topology and establish if they are an end-device, a fog-device, or neither.
These associations are key to differentiating devices and establishing what actions are appropriate to perform on a particular device.
For example, the FDK only instantiates services on fog-devices, as such an action would not be appropriate for other devices.
Section \ref{ResourceManager} presents this mechanism in detail.





\subsection{FlowManager}
\label{FlowManager}
The \textit{FlowManager} component provides a comprehensive interface for the management of OpenFlow flows throughout the network.
The core APIs for this component are described in Table \ref{tab:flowAPIs}.
\begin{table}[t]
    \centering
    \caption{FlowManager APIs}
    \begin{tabular}{|l|p{5cm}|}
    \hline
    \textbf{API} & \textbf{Description} \\ \hline \hline
    \texttt{\scriptsize{create\_flow()}} & Push OpenFlow flow to switch\\ \hline
    \texttt{\scriptsize{delete\_flow()}} & Delete OpenFlow flow from switch \\ \hline
    \texttt{\scriptsize{track\_flow()}} & Track flow information \\ \hline
    \texttt{\scriptsize{untrack\_flow()}} & Untrack flow information \\ \hline
    \end{tabular}
    \label{tab:flowAPIs}
\end{table}
First, the FlowManager provides a set of APIs to simplify the process of creating flow table entries on OpenFlow switches.
For example, this component provides a method for creating a \textit{flow skeleton}, which contains all of the basic fields needed to create the flow table entries used by the FDK to enforce traffic paths between end-devices and fog-devices.
Then, the FlowManager's flow-modification APIs can be utilized to further build and shape entries by adding flow actions, flow match fields, and other constructs to a flow skeleton.
For example, flows can be created to match packets by source and destination IP address (or additional identifiers).
Upon a match, multiple actions can be applied to a packet---such as transmitting it through a specific port (used to create network traffic paths) and placing it on a packet queue.
Once a flow table entry is built, the FlowManager's flow-creation APIs can be leveraged to push a newly-built entry to an OpenFlow switch. Similarly, the FlowManager offers flow-deletion APIs that can be used to remove such entries.

%
%
%
%
%
%

\subsection{ResourceManager}
\label{ResourceManager}
The FDK uses the \textit{ResourceManager} component to manage and allocate all networking and computing resources.
The core APIs of this component are described in Table \ref{tab:resourceAPIs}.
\begin{table}[t]
    \centering
    \caption{ResourceManager APIs}
    \begin{tabular}{|l|p{3.75cm}|}
    \hline
    \textbf{API} & \textbf{Description} \\ \hline \hline
    \texttt{\scriptsize{service\_end\_device()}} & Process service requests from end-devices, run the RAA, and instantiate containers \\ \hline
    \texttt{\scriptsize{service\_shutdown\_request()}} & Process shutdown requests, run the RDA, and shutdown containers \\ \hline
    \texttt{\scriptsize{service\_fog\_device()}} & Receive and process resource reporting messages from fog-devices \\ \hline
    \texttt{\scriptsize{resource\_alloc\_algorithm()}} & Attempt to allocate all resources for requested service \\ \hline
    \texttt{\scriptsize{resource\_dealloc\_algorithm()}} & Attempt to deallocate all resources for a service \\ \hline
    \end{tabular}
    \label{tab:resourceAPIs}
\end{table}
The ResourceManager maintains data structures regarding all resources available in the network.
This is possible with the help of an agent running on every fog-device. 
This agent continually collects and relays information (such as processor and memory utilization) back to the ResourceManager over time.
Similarly, the ResourceManager also repeatedly queries the ODL node inventory to gather current link utilization information.
This information is then stored in the Topology data structure managed by the TopologyManager, which ultimately provides a complete overview of all available resources throughout the network.

The main functionalities provided by the ResourceManager lie within the servers that enable the end-devices to request/release computing resources.
These servers act as an interface for managing containerized services and the allocation of resources in the fog-devices.
For example, the \textit{service request server} receives and processes requests from end-devices, where each request specifies parameters such as an image name of a containerized service to run and a set of resource requirements for the request.
The image name refers to the type of application processing requested. 
For example, an end-device may specify an image implementing a medical classification application.

Once a request is received, the ResourceManager executes the \textit{resource allocation algorithm} (RAA) presented in Algorithm \ref{RAA}.
If sufficient resources exist, the desired containerized service with the appropriate amount of resources is instantiated on a fog-device, a communication path between the end-device and the fog-device is reserved, and a bandwidth allocation along that path is enforced.
Conversely, the \textit{shutdown request server} provides an interface to revert this process by shutting down containers and deallocating resources.

\SetEndCharOfAlgoLine{}
\begin{algorithm}[!ht]
\label{RAA}
\footnotesize
\KwIn{\\
$e_{i} = $ end-device requesting resources\\
$R_{B}(e_{i}) = $ Bandwidth requirement of request from $e_{i}$\\
$R_{P}(e_{i}) = $ Processing requirement of request from $e_{i}$\\
$R_{M}(e_{i}) = $ Memory requirement of request from $e_{i}$\\
Complete topology and resource data (from TopologyManager)\\
}

\BlankLine

\KwOut{\\
A response for $e_{i}$ indicating success or failure\\
}
\BlankLine
\BlankLine
$T_{B}(l) = $ Total bandwidth capacity on link $l$ \label{band_all_def_start}\\
$T_{P}(f) =$ Total processing capacity on fog-device $f$\\
$T_{M}(f) =$ Total memory capacity on fog-device $f$\\

\BlankLine
$A_{B}(l) = $ Allocated bandwidth on link $l$\\
$A_{P}(f) =$ Allocated processing on fog-device $f$\\
$A_{M}(f) =$ Allocated memory on fog-device $f$ \label{band_all_def_end} \\

\BlankLine

$\mathcal{N} = $ Set of all nodes \\
$\mathcal{L} = $ Set of all links \\
$\mathcal{F} = $ Set of all fog-devices \\
$\mathcal{F}^{\prime} = \emptyset$ \texttt{//Request servicers} \\
$\mathcal{P} = \emptyset$ \texttt{//Shortest-path tree} \\
$\mathcal{B} = \emptyset$ \texttt{//Best known link dictionary} \\

\BlankLine
\texttt{//identify request servicers} \\
\For {$f_{j} \in \mathcal{F}$}
{
    \label{raa:createRequestServicers}
    \If {$T_{P}(f_{j}) - A_{P}(f_{j}) > R_{P}(e_{i}) \And$ \\
          $T_{M}(f_{j}) - A_{M}(f_{j}) > R_{M}(e_{i})$} 
    {
        Add $f_{j}$ to $\mathcal{F}^{\prime}$ \\ 
    }
}

\BlankLine

\lIf {$\mathcal{F}^{\prime} == \emptyset$}
{
    return FAILURE response \label{raa:failureResponse1}
}

\BlankLine
\label{raa_set_of_nodes}
$k = \mathrm{max}(2, \mathrm{size}(\mathcal{L})/\mathrm{size}(\mathcal{N}))$ \\
$\mathcal{H} = minHeap(k)$ \texttt{//K-ary min heap}\\
\BlankLine

\label{raa:dummy_edge}
$init\_link = (src: e_{i}, dst: e_{i}, weight: 0)$ \\
$\mathcal{H}.\mathrm{push}(init\_link)$ \\
$\mathcal{B}[e_{i}] = init\_link$ \\

\BlankLine
\texttt{//find least-cost paths from $e_{i}$ to fog-devices} \\
\While {$\mathrm{size}(\mathcal{H}) > 0$} 
{
    $u = \mathcal{H}.\mathrm{pop\_min()}$\\
    
    \If{$u.src \neq u.dst$} {
        $\mathcal{P}[u.dst] = u$ \label{RAAshortestpathadd}
    }
    
    \For {$v \in \{ \mathrm{outgoing\;links\;of\;} u.dst\}$}
    {
        \texttt{//v.src is equivalent to u.dst}\\
        $v.weight = \mathcal{B}[v.src].weight + 1/(T_{B}(v) - A_{B}(v))$ \label{RAAweight} \\
        \lIf {$(T_{B}(v) - A_{B}(v)) < R_{B}(e_{i})$} {$v.weight = \infty$}
        
        \If{$v.dst \not\in \mathcal{B}$} {
            $\mathcal{H}.\mathrm{push}(v)$ \label{RAApush}\\
            $\mathcal{B}[v.dst] = v$
        }
        \ElseIf{$v.weight < \mathcal{B}[v.dst].weight$} {
            \texttt{//Update link, shift based on weight}\\
            $\mathcal{H}.\mathrm{decrease\_key}(\mathcal{B}[v.dst], v)$ \label{RAAdecreasekey} \\
            $\mathcal{B}[v.dst] = v$
        }
    }
}

\BlankLine

\texttt{//find the best fog-device to fulfill the request} \\
$min = \infty$ \\
\For {$f_{j} \in \mathcal{F}^{\prime}$}
{
    \If {$\mathcal{P}[f_{j}].weight < min$}
    {
        $min = \mathcal{P}[f_{j}].weight$ \\ \label{raa:trackMin}
        $f_{min} = f_{j}$ \label{raa:selectFog}
    }
}

\lIf {$min == \infty$}
{
    return FAILURE response \label{raa:failureResponse2}
}

\BlankLine

\texttt{{//configure switches along the path $f_{min}$ to $e_{i}$}} \\
$v = \mathcal{P}[f_{min}]$\\
\While {true} 
{
    \lIf {$v.src == e_{i}$} 
    {
        return SUCCESS response \label{raa:responseMsg}
    }
    
    Create rate-limited queues on $v.src$\\ \label{raa:createQueue}
    Place queues on appropriate QoS entry in $v.src$\\ \label{raa:placeQueue}
    Create flows on $v.src$ to redirect traffic to rate-limited queues\\ \label{raa:createFlow}
    
    $v$ = $\mathcal{P}[v.src]$
}

\caption{Resource Allocation Algorithm (RAA)}
\label{RAA}
\end{algorithm}

The RAA uses a modified version of Dijkstra's shortest-path algorithm in addition to some pre- and post-processing steps.
The implementation of Dijkstra's algorithm leverages a $k$-ary min heap for optimal real-world performance \cite{shortest_path_algs_eval}.
If $n$ is the number of nodes and $m$ is the number of links, then $k=\mathrm{max}(2, m/n)$ is the number of children per node in the $k$-ary heap. 
It has been shown that this algorithm has a run-time complexity of $O(m\log_{k}n)$\cite{sedgewick_2002}.
Although there are theoretically faster implementations of this algorithm using a Fibonacci heap, the $k$-ary heap implementation is known to be significantly faster in real-world scenarios \cite{shortest_path_algs_eval}.
The RAA's inputs are an end-device $e_{i}$, the resources requested by $e_{i}$, and complete topology data.

The RAA begins with a pre-processing step, where it iterates over all fog-devices $f_{j}$ and assesses their available resources to create a list of \textit{request servicers} $\mathcal{F'}$  (line \ref{raa:createRequestServicers}).
Specifically, $\mathcal{F'}$ is a list of fog-devices that have sufficient resources to fulfill the request.
Afterwards, if no request servicers exist, then the RAA returns a failure response that is subsequently sent back to $e_{i}$ by the ResourceManager (line \ref{raa:failureResponse1}).

If at least one request servicer exists, then the RAA continues and executes Dijkstra's shortest-path algorithm to find the shortest path from $e_{i}$ to all other nodes in the topology.
The algorithm defines the cost across any link $l$, from the node $l.src$ to the node $l.dst$, as $1/(T_{B}(l) - A_{B}(l))$.
It is important to note that the amount of available bandwidth on the link $l$, computed as $T_{B}(l) - A_{B}(l)$, is never affected by control (and normal background) traffic (e.g., OVSDB messages, service requests, etc.) because a separate allocation for this traffic is made when the FDK initially starts.
However, the actual weight of $l$ is defined as the total cost required to reach $l.dst$ from $e_{i}$ (unless there is insufficient bandwidth on $l$ to fulfill the request, in which case the weight is $\infty$).
To this end, the algorithm uses a dictionary $\mathcal{B}$ which tracks the best known links used to reach nodes from $e_{i}$.
Therefore, we say that $l.weight = \mathcal{B}[l.src].weight + 1/(T_{B}(l) - A_{B}(l))$, where $B[l.src]$ is the best known link used to reach $l.src$ from $e_{i}$ (line \ref{RAAweight}).
Using this weight relies on the fact that links are stored on a $k$-ary min heap $\mathcal{H}$, which then keeps the link with the lowest weight at the top.
This implies that any link $l$ at the top of the heap can be used to reach $l.dst$ with the lowest possible total cost from $e_{i}$ (assuming $l.src \neq l.dst$), meaning $l$ is suitable to be added to the shortest-path tree $\mathcal{P}$.
This link weight also results in the selection of paths throughout the network which tend to be short and have a high amount of available bandwidth.
Furthermore, $\mathcal{H}$ is continually updated during algorithm execution with the help of the best known link dictionary $\mathcal{B}$.
Specifically, $\mathcal{B}[n]$ returns the best known link to reach node $n$.
If $n \not \in \mathcal{B}$ ($n$ has not been reached already), then the link used to reach $n$ is pushed onto $\mathcal{H}$ (line \ref{RAApush}).
Otherwise, $n$ has been reached already and the RAA checks if the new link used to reach $n$ has a lower weight than the best known link $\mathcal{B}[n]$.
If it does, then a modified decrease-key operation is performed on $\mathcal{H}$ which replaces the link $\mathcal{B}[n]$ with the cheaper new link (line \ref{RAAdecreasekey}).
Then, the new link is shifted upwards in the heap.
This process repeats as the algorithm continues to visit nodes using different paths, eventually shifting the best links to the top of $\mathcal{H}$ and choosing to include them in the shortest-path tree dictionary $\mathcal{P}$ (line \ref{RAAshortestpathadd}).

Once Dijkstra's algorithm is finished, dictionary $\mathcal{P}$ contains the shortest-path tree.
To be more precise, $\mathcal{P}[n]$ returns the link attached to $n$ facing $e_{i}$ that is included in the shortest path from $e_{i}$ to $n$, as well as its weight and both nodes at the endpoints of the link.
To this end, $\mathcal{P}$ can be used to traverse and gather information on the shortest path between $e_{i}$ and any other device in the network.

$\mathcal{P}$ is  then used in the subsequent post-processing step. 
First, $\mathcal{P}[f_{j}].weight$ is checked for all $f_{j} \in \mathcal{F'}$ and a fog-device $f_{min} \in \mathcal{F'}$, where  $\mathcal{P}[f_{min}].weight = \mathrm{min}(\mathcal{P}[f_{j}].weight) \; \forall\;f_{j}\in\mathcal{F'}$ is selected to fulfill the service request (line \ref{raa:selectFog}). 
If $\mathcal{P}[f_{min}].weight$ = $\infty$, then no paths with sufficient bandwidth between $e_{i}$ and any request servicers exist, and a failure response is returned to $e_{i}$ as a result (line \ref{raa:failureResponse2}).
Otherwise, the path between fog-device $f_{min}$ and end-device $e_{i}$ has a sufficient amount of bandwidth and $f_{min}$ is chosen to fulfill the request from $e_{i}$.

The next step is to allocate network resources along the identified path between $e_{i}$ and $f_{min}$.
The nodes along this path are accessed by traversing through dictionary $\mathcal{P}$.
Network resource allocation begins with the creation of \textit{rate-limited} queues on each switch along this path.
The ResourceManager accomplishes this by making a call to the TopologyManager function \texttt{create\_queue()}, which leverages the OVSDB management protocol to create and configure the queues (line \ref{raa:createQueue}).
The rate-limit is specified in the queue configuration data and is equal to $R_{B}(e_{i})$.
Once created, these queues are placed on QoS entries (created on startup of the FDK by the TopologyManager) using a similar TopologyManager function \texttt{place\_queue\_on\_qos()} (line \ref{raa:placeQueue}).
These QoS entries map to switch ports connected to the network links along this path, effectively resulting in each port having a set of packet queues that limit egress traffic.

\begin{figure}[t]
\centering
\includegraphics[width=1\linewidth]{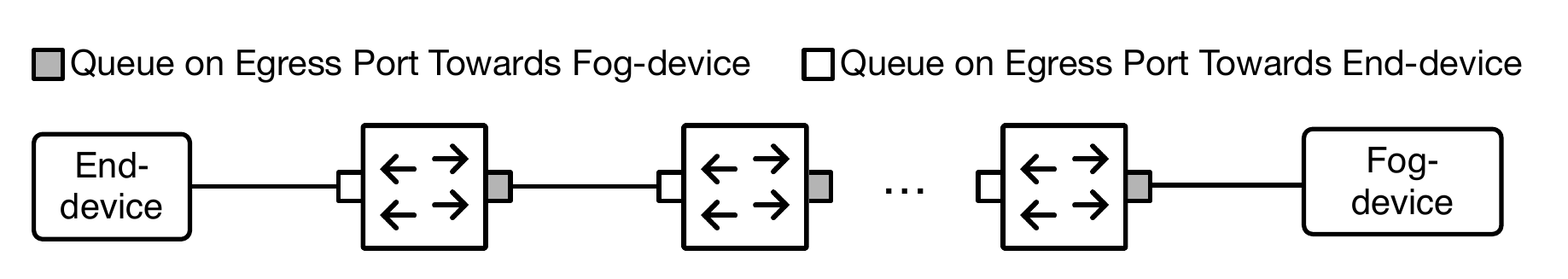}
  \captionof{figure}{Enforcing bandwidth reservation using rate-limited queues. 
  For each path reservation, rate-limited packet queues are created and attached to QoS configurations located on the egress ports towards the fog-device as well as those towards the end-device. Then, flow table entries are pushed via OpenFlow to enqueue traffic traveling from the end-device to the fog-device, and vice versa, on these queues.
  }
  \label{fig:linkReservation}
\end{figure}

In addition to queues, flows must also be created to ensure that traffic is directed along the identified path between $e_{i}$ and $f_{min}$ and that packets exchanged between the two devices are placed on the proper queues within each switch.
Therefore, as the ResourceManager installs packet queues on each switch, it also uses OpenFlow to redirect traffic along the identified path and to the appropriate queues along that path by leveraging the FlowManager flow-creation APIs (line \ref{raa:createFlow}).
Each OpenFlow flow specifies a set of actions for the reserved path. 
Therefore, on each switch along the path the FDK uses one OpenFlow flow that specifies multiple actions: one for redirecting traffic to the desired port (therefore reserving a one-way path for communications between $e_{i}$ and $f_{min}$), and another to place packets on the appropriate queue for that port.
Similarly, for communications in the opposite direction from $f_{min}$ to $e_{i}$, another packet queue and OpenFlow flow is installed on each switch.
Therefore, the overhead of enforcing a path and reserving communication bandwidth for one service involves the creation of two packet queues and two OpenFlow flows on each switch along the identified path.
Figure \ref{fig:linkReservation} depicts the creation of rate-limited queues along a path to ensure network bandwidth allocation in both directions.
Finally, because the FDK never over-allocates resources, the rate-limiting of bandwidth effectively results in the \textit{allocation} of bandwidth.

The flows installed on switches match packets (flow classification) based on source IP address, destination IP address, source or destination port number (depending on the traffic direction), and protocol type.
For communications from $e_{i}$ to $f_{min}$, the source IP address is $e_{i}$'s IP address, the destination IP address is $f_{min}$'s IP address, and the destination port is a proxy port on $f_{min}$ assigned to the containerized service.
The protocol type specifies the transport layer protocol used by the application.
The transport layer protocols supported are UDP, TCP, SCTP, and any user-space protocol that relies on these protocols.
For example, QUIC \cite{quic1,quic2} is a widely-used user-space protocol that is implemented on top of UDP, and is therefore supported by the FDK.

Finally, a success response containing $f_{min}$'s IP address and the proxy port (if the end-device has not asked for a particular port number) is returned to the service request server (line \ref{raa:responseMsg}), which then remotely instantiates a container on $f_{min}$ using Docker Swarm. 
The success response is then forwarded to the end-device as well.
At this point, all computational and networking resources have been allocated, and once $e_{i}$ receives the success response message, it can begin communicating with the newly created containerized service running on $f_{min}$.
%
%

There are multiple mechanisms available to direct packets to the appropriate container when they arrive at $f_{min}$.
The first mechanism is to dedicate a unique proxy port on the fog-device to each service. 
To this end, for each containerized service, the FDK finds a unique port number that has not been used on the fog-device hosting the container.
The FDK also allows end-devices to specify their desired destination port number when making requests. 
However, without adding additional capabilities, this mechanism does not allow two or more end-devices to request the same port number on a fog-device.
To address this issue, the fog-device demultiplexes (using reverse proxy or OVS) the received packets to different containers based on their source IP address.
Therefore, once a service request is fulfilled, the FDK only needs to return the IP address of the identified fog-device to the end-device.
An alternative approach to supporting multiple containers using the same port numbers on the same fog-device is to assign each container an IP address in the same subnet as that of the fog-device.
In this case, the IP address assigned to the container is returned to the end-device, instead of the IP address of the fog-device.
Also, the container's IP address is used to configure the flow tables on switches along the communication path.
This approach, however, is not officially supported by Docker due to its security issues.
Specifically, this approach does not allow the protection of containers from the outside world and from each other.
In contrast, using a proxy port requires ingress access to be explicitly granted, which offers higher security.
Therefore, although both mechanisms are supported by the FDK, in this work we particularly focused on the former due to its higher security and wide-spread adoption \cite{dockerNetworking}.

As seen throughout this section, applications must be adapted to the FDK in order to benefit from fog resources.
Therefore, end-devices must be programmed to issue service requests so that these resources may be allocated.
However, this may not be possible with commercial, non-open-source applications running on the end-devices.
A simple solution is to use a middleware that issues service requests on behalf of the application. 
Also, since the middleware can translate the destination port number of packets originating from end-devices, a unique port number can be assigned to each request, and therefore there is no need to use a demultiplexing tool on the fog-devices to deliver incoming packets to the appropriate container.
It is also worth noting that the middleware does not need to be implemented on the end-devices.
As an example, consider a gateway node (such as a smartphone or an IoT gateway) collecting data from multiple sensing devices.
The gateway can then request for resources on behalf of these devices, and therefore there is no need to modify the software stack of the sensing devices.

To summarize, consider a scenario where multiple end-devices communicate with multiple containers that run on a single fog-device and listen on the same port.
In this case, source IP address is the 5-tuple's element that is used to classify these flows by the switches as well as the fog-device.
Alternatively, if a gateway that includes a middleware is used to issue requests on behalf of multiple end-devices, since the source IP address of all the requests generated by the gateway are the same, the FDK generates a unique port number assigned to each service.
In this case, port number is the 5-tuple's element that is used to classify these flows by the switches as well as the fog-device.
Therefore, the FDK offers a robust flow classification mechanism on switches and fog-devices as a part of its networking and computational resources.
With this mechanism, end-devices are provided with the capability to make multiple service requests in parallel.
This implies that any end-device may request an arbitrary amount of services (as long as sufficient resources exist), and therefore run an arbitrary number of fog applications.

As the ResourceManager continues allocating resources over time, it keeps track of all allocated resources.
Once an end-device decides to terminate a service, it issues a shutdown request to the \textit{shutdown request server}, which then runs the \textit{resource deallocation algorithm} (RDA).
The RDA identifies and releases the resources allocated for the corresponding service.
In short, OpenFlow flows along the reserved path are deleted, network bandwidth is deallocated by deleting the appropriate packet queues, and the containerized service in the fog-device is shutdown.

\section{Evaluation}
\label{Evaluation}


In this section we first verify the correctness and performance of the FDK using sample applications running on a physical testbed.
We then present simulation-based scalability analysis of the FDK.




%
%
%

\subsection{Verification and Evaluation using a Physical Testbed}
\label{phy_eval}
In this section we verify the correctness and performance of the FDK using a physical testbed running various applications.

\begin{figure}[t]
\centering
\includegraphics[width=0.95\linewidth]{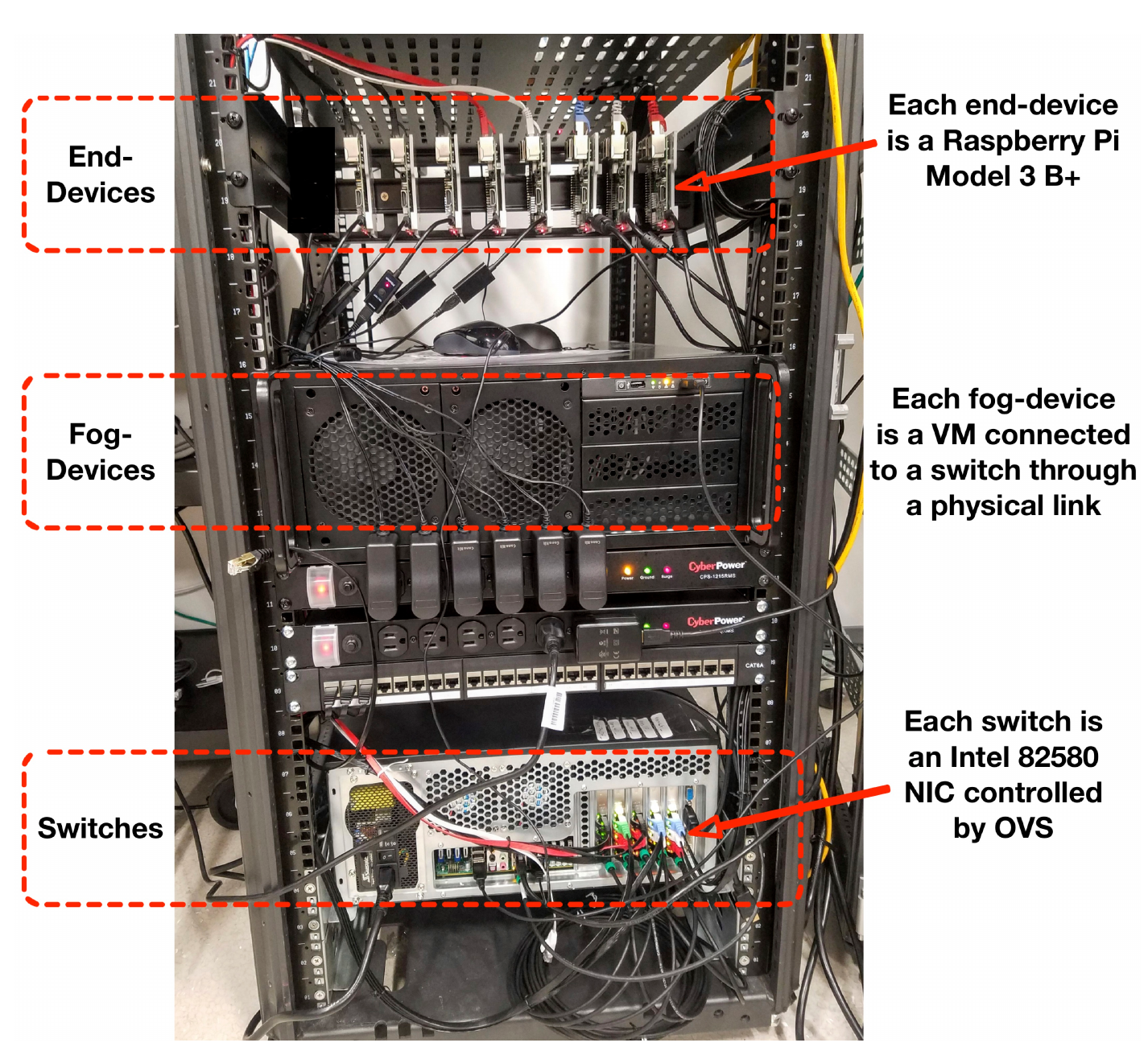}
  \captionof{figure}{The physical testbed used to implement the topology depicted in Figure \ref{topology}.
  End-devices, switches, and fog-devices are connected through physical links.}
  \label{physicalInfra}
\end{figure}

\begin{figure}[t]
\centering
\includegraphics[width=1.0\linewidth]{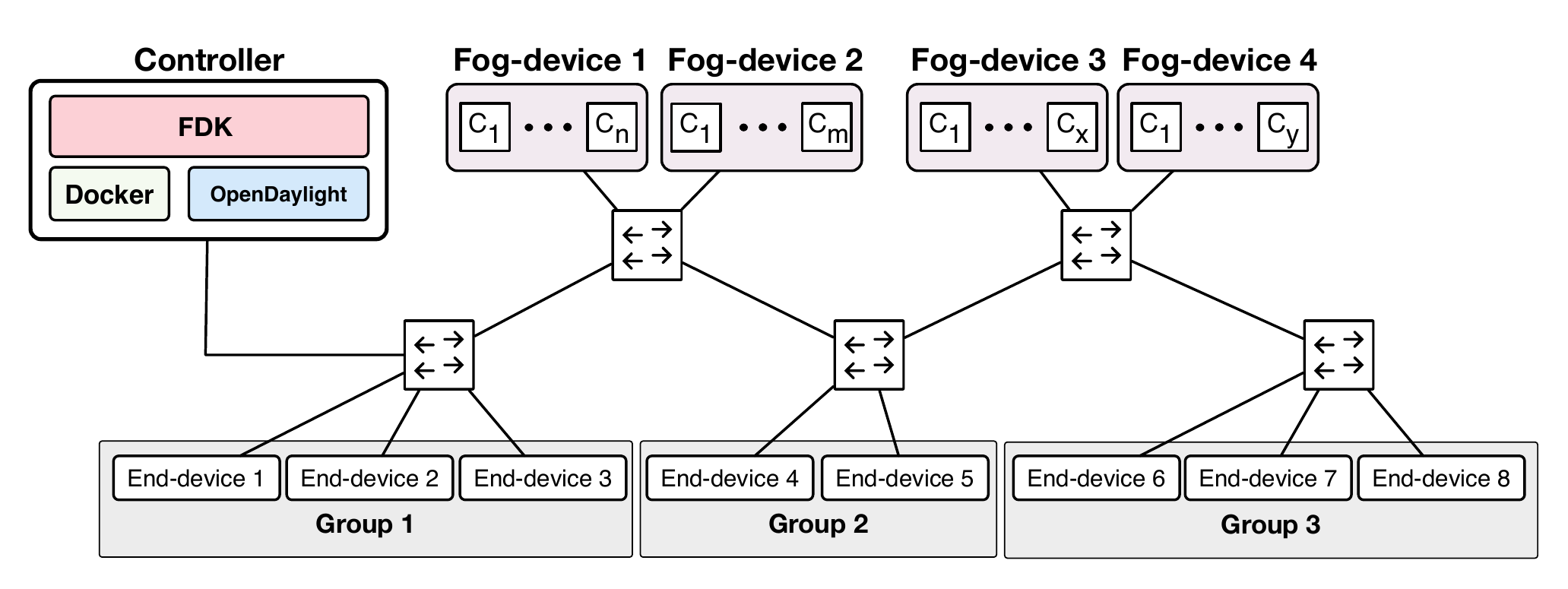}
  \captionof{figure}{The network topology used for the development and testing of the FDK. $C_i$ represents a container running on a fog-device.}
  \label{topology}
\end{figure}

\textbf{Testbed.} Figure \ref{physicalInfra} shows our testbed, which includes five OpenFlow switches, four fog-devices, and eight end-devices.
This testbed implements the network presented in Figure \ref{topology}.
Each end-device is a Raspberry Pi Model 3 B+ (running Raspbian Linux) which is connected to a switch using a 1 Gbps cable.
The machine hosting the four fog-devices includes a 4-port Intel 82580 NIC, where each fog-device is a VM associated with a physical port.
Another machine includes five 4-port Intel 82580 NICs as well as a 2-port NIC to build the five OpenFlow switches.
The 2-port NIC is paired with one of the aforementioned 4-port NICs to build a 6-port switch which is connected using a 1 Gbps cable to the controller.
Both machines include two 16-core Intel Xeon CPUs and 64 GB RAM.
Each fog-device and OpenFlow switch uses Ubuntu Server 18.10 and leverages 4 CPU cores and 8 GB of RAM.
The OpenFlow switches run OVS 2.10.0 and support both OpenFlow 1.3 and OVSDB.
Docker daemons run on each fog-device, and are configured to listen for remote TCP connections from the controller.
The controller (including the FDK) is hosted on an external server. 

%
%

We partitioned the end-devices into three groups.
Referring back to Figure \ref{topology}, we placed end-devices 1, 2, and 3 into \textit{Group 1}, end-devices 4 and 5 into \textit{Group 2}, and end-devices 6, 7, and 8 into \textit{Group 3}. 
This grouping helps us identify the effect of the FDK on network overhead, which may vary depending on the location of the end-devices.
For example, assume that all the end-devices issue service requests concurrently.
The OpenFlow switch connected to the devices in Group 1, which is the switch closest to the controller, would be placed under higher stress compared to those switches further from the controller, such as the switch connected to Group 3. 
In this case, all service and shutdown request messages, OpenFlow messages, OVSDB messages, Docker Swarm container instantiation messages, etc., pass through the switch connected to Group 1.
At the same time, only a fraction of these messages passes through the switch connected to Group 3.
We created the three Groups in an attempt to capture the effect of these variations.


\textbf{Applications.} We evaluate the \textit{application development capabilities} of the FDK by creating a set of sample applications.
The first application, which includes an iperf3 server and an iperf3 client, is called \textit{iperf-app} and enables an end-device (client) to communicate through TCP with a containerized service on a fog-device (server).
To develop \textit{iperf-app}, we first created a Python script that hosts an iperf3 server using the iperf-python library \cite{iperf-python}.
We then packaged this script into a Docker image.   
Finally, we modified the server script to communicate all bandwidth readings to a background process running on each fog-device. 
This process receives and saves the readings.
On the end-devices, we created another Python application that issues a service request to instantiate the aforementioned Docker image as a container, starts the client that streams data to the server running in the container, and then issues a shutdown request once the client terminates.
The second application developed is \textit{sleep-app}, which sends a service request, sleeps for a particular duration, and then sends a shutdown request.
These applications are used to analyze the impact of service requests and varying levels of bandwidth utilization on the FDK's ability to service those requests.
The third application developed is an object detection application named \textit{detection-app}.
The application streams image data from end-devices to the services in the fog-devices, which run object detection algorithms to identify different objects found in images.
The transport protocol used by this application is QUIC.
A real-world example of this application is an object classification and packaging system.
Another application is a real-time surveillance system supporting facial recognition.

\begin{figure}[t]
\centering
\includegraphics[width=0.97\linewidth]{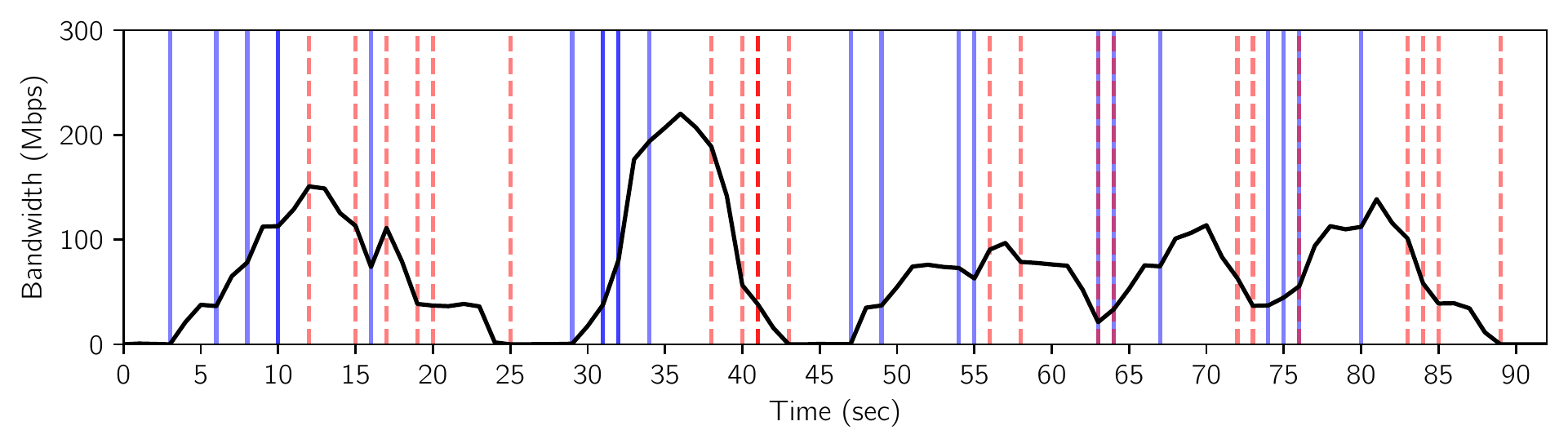}
  \captionof{figure}{Overall bandwidth of the data received by fog-devices corresponding to the images captured and sent by end-devices via \textit{detection-app}. 
  Blue (solid) and red (dashed) bars denote service requests and shutdown requests made by end-devices, respectively.
  Bars which are less transparent indicate a greater amount of service or shutdown requests made during a particular second.
  This figure shows the dynamics of allocating and deallocating resources by the FDK when end-devices randomly issue service and shutdown requests.
  }
  \label{quic_app}
\end{figure}

\textbf{Verification.} Before running any tests, and in order to confirm the functionality of the FDK, we issued service requests to the FDK from the end-devices and verified that resources are allocated properly.
To this end, we made temporary modifications to the fog-side Docker images that would consume as many resources as possible and then confirmed that the containers instantiated from these images did not exceed the resources allocated to them.
For example, we modified \textit{iperf-app} in one test to spin up an infinite \texttt{while} loop script that consumed all processing resources.
Then, by using performance monitoring tools such as \texttt{top} we confirmed that the container did not exceed the resource allocations requested by the end-device.
Similarly, we confirmed that network resources were appropriately allocated using \textit{iperf-app}, which revealed that bandwidth allocations were not exceeded.
Finally, we used \textit{detection-app} to represent a real-world scenario, where configured end-devices randomly wake-up, issue a service request, and then capture and stream images to the fog-devices running object recognition algorithms. 
Each service request specifies a desired bandwidth allocation of 40 Mbps. 
Each end-device, after about 7 seconds into its streaming period, ceases its image streaming and sends a shutdown request.
We performed similar verification steps to ensure that resources were all allocated and deallocated properly.
Figure \ref{quic_app} shows the total bandwidth of streaming data received by the fog-devices.
To generate this figure, all fog-devices were configured to be time-synchronized, and each container was configured to record the number of bytes received per second through its network interface.
%
%

Given that these applications use a variety of transmission rates, transport layer protocols, and randomized service and shutdown request patterns, the operation of the FDK was carefully verified before proceeding with performance evaluation tests.
In the rest of this section, we present performance evaluation of the FDK.


%
%
%
\noindent
\subsubsection{Test 1: Resource Allocation and Deallocation}
The goal of Test 1 is to characterize the computational and communication overhead of the FDK.
This is accomplished by running applications across all end-devices and recording the runtimes of various operations under different circumstances.
We track the duration of key operations including resource allocation (RAA), resource deallocation (RDA), service request fullfillment, and shutdown request fulfillment.
Service request fulfillment duration refers to the total duration between the time an end-device sends a service request to the FDK and the time the end-device receives a success response from the FDK.
Similarly, shutdown request fulfillment duration refers to the total duration between sending a shutdown request and the reception of confirmation.
For this experiment, we ran \textit{sleep-app} across all eight end-devices in the topology and measured the duration of the aforementioned performance parameters.
We repeated this experiment 250 times for a total of 2000 \textit{sleep-app} runs, and ran two different versions of this test, bringing the number to 4000.
These different test versions are \textit{Test 1a} and \textit{Test 1b}, as follows.

\begin{figure*}[t]
\centering
\includegraphics[width=1\linewidth]{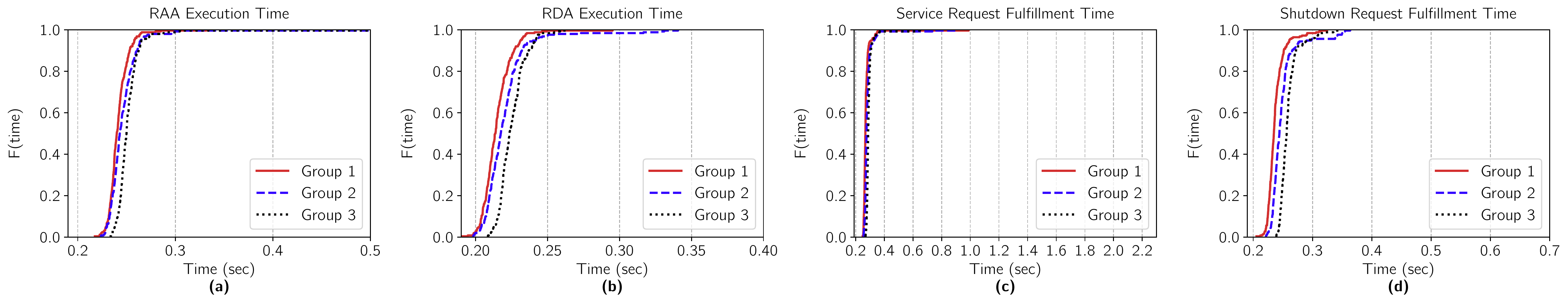}
  \captionof{figure}{Empirical Cumulative Distribution Function (ECDF) graphs for Test 1a. 
  In this test, end-devices issue service requests \textit{sequentially}. 
  Groups closer to the controller (and therefore FDK) complete all of their operations slightly faster than those further from the controller.
  }
  \label{test1a}
\end{figure*}

\textbf{Test 1a.} In this test, the end-devices \textit{sequentially} run \textit{sleep-app}.
For example, end-device 1 issues a service request, sleeps for 3 seconds after receiving service, and then issues a shutdown request. 
After completion, the rest of the end-devices perform the same operation sequentially.
Figure \ref{test1a} presents the results of Test 1a.
The duration of various operations are averaged out among the end-devices of each Group and are then displayed as ECDF graphs.
As seen in Figure \ref{test1a}, more than 95\% of all operations completed within 0.33 seconds across all Groups.
In addition, resource allocation times and service request fulfillment times are nearly identical, as are the resource deallocation times and shutdown request fulfillment times.
This means that resource allocation is the main source of overhead in the process of fulfilling service requests, and that resource deallocation is the main source of overhead in the process of fulfilling shutdown requests.
Also, operations performed for devices in Group 1 tend to finish slightly faster than those for Group 2, which finish faster than those for Group 3.
This is caused by the shorter queuing and packet processing delays along the path to the controller with a fewer number of switches.
However, the difference in timing is on the order of a few milliseconds.

\begin{figure*}[t]
\centering
\includegraphics[width=1\linewidth]{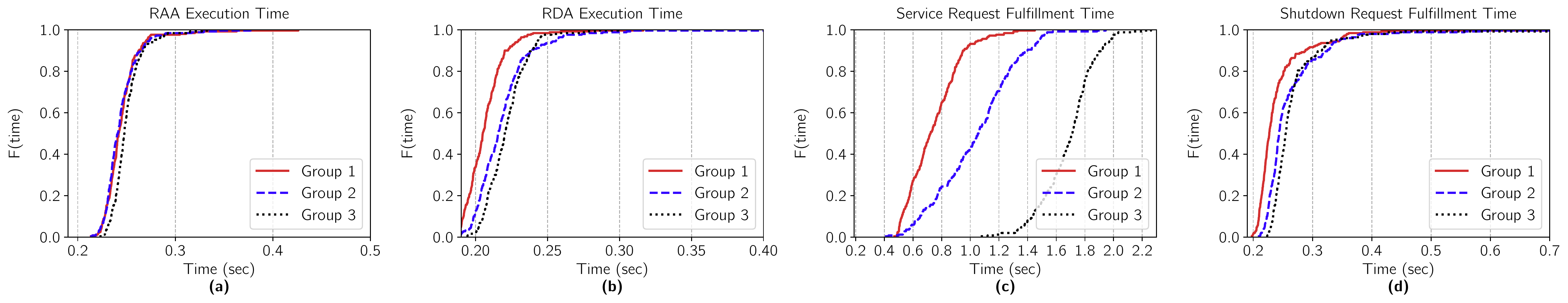}
  \captionof{figure}{Empirical Cumulative Distribution Function (ECDF) graphs for Test 1b. 
  In this test, end-devices issue service requests \textit{concurrently}. 
  Groups closer to the controller experience significantly faster service request fulfillment times compared to those further from the controller.
  This is because the FDK processes requests \textit{sequentially}. 
   }
  \label{test1b}
\end{figure*}

\textbf{Test 1b.} In this test, the end-devices \textit{concurrently} run \textit{sleep-app}.
In this case, all end-devices issue a service request to the FDK at the same time, sleep for 3 seconds upon receiving a successful response, and then send a shutdown request.
Figure \ref{test1b} presents the results of Test 1b.
The results presented in Figures \ref{test1b}(a) and \ref{test1b}(b) are nearly identical to the corresponding Figures \ref{test1a}(a) and \ref{test1a}(b) from Test 1a, with 95\% of these operations completing within 0.28 seconds across all Groups.
However, the results for service request fulfillment times in Test 1b, shown in Figure \ref{test1b}(c), look considerably different compared to the corresponding Figure \ref{test1a}(c) from Test 1a.
Here, we can observe greater variations in the results, with Group 1, Group 2, and Group 3 showing median service request fulfillment durations of 0.72, 1.06, and 1.71 seconds, respectively.
Also, there is far less variation in the results for shutdown request fulfillment times, as Figure \ref{test1b}(d) shows.
In this regard, Group 1, Group 2, and Group 3 show median shutdown request fulfillment durations of 0.23, 0.24, and 0.25 seconds, respectively.

Because the service fulfillment process accesses and modifies various shared data structures such as the Topology object representing the current state of the network, the entire process is guarded by a mutex.
This means that the FDK queues concurrent service requests and handles them sequentially.
This effect can be seen in Figure \ref{test1b}(c).
Here, because the end-devices in Group 1 are closer to the controller than those in Groups 2 and 3, service requests from these devices sit closer to the front of the queue than the requests arriving later from Groups 2 and 3.
Therefore, Groups 2 and 3 experience slower service request fulfillment times compared to Group 1.
Similarly, the process of resource deallocation is also guarded by a mutex, meaning that concurrent shutdown requests are handled sequentially as well.
However, because the service requests are fulfilled sequentially, the sleep durations and subsequent shutdown requests made by each \textit{sleep-app} instance become desynchronized and happen sequentially.
As a result, we see a much smaller impact on shutdown request fulfillment times in comparison to service request fulfillment times in Test 1b.

\subsubsection{Test 2: Bandwidth Guarantee}
In Test 2, we evaluate the overhead of the FDK on the network.
Specifically, we investigate if the FDK compromises bandwidth allocations (by reducing transmission speeds) for running fog applications.
We chose one end-device from each Group to run \textit{iperf-app} for 90 seconds with a 300 Mbps bandwidth allocation. 
This is the maximum transmission rate of the Raspberry Pi Model 3 B+.
Also, after subtracting transmission overheads such as packet headers, the actual data transmission rate supported is around 280 Mbps.
Using \textit{iperf-app}, an end-device continuously streams data to a container for 90 seconds.
Then, at 30 and 60 seconds into the 90-second transmission, all 7 other end-devices in the topology run \textit{sleep-app} for one second.
This results in a group of service requests, shutdown requests, OpenFlow messages, OVSDB messages, and Docker Swarm container instantiation messages flowing through the network.
We ran this experiment 100 times for the chosen end-device and repeated it for the other two chosen end-devices from the other two Groups, for a total of 300 \textit{iperf-app} runs and 4200 \textit{sleep-app} runs.
Finally, we used two separate versions of this test and analyzed their impact on network congestion and the transmission bandwidth of \textit{iperf-app}.
In the end, 600 \textit{iperf-app} runs and 8400 \textit{sleep-app} runs were performed.
The two modified test cases, called Test 2a and Test 2b, are outlined in detail as follows.

\begin{figure}[t]
\centering
\includegraphics[width=0.84\linewidth]{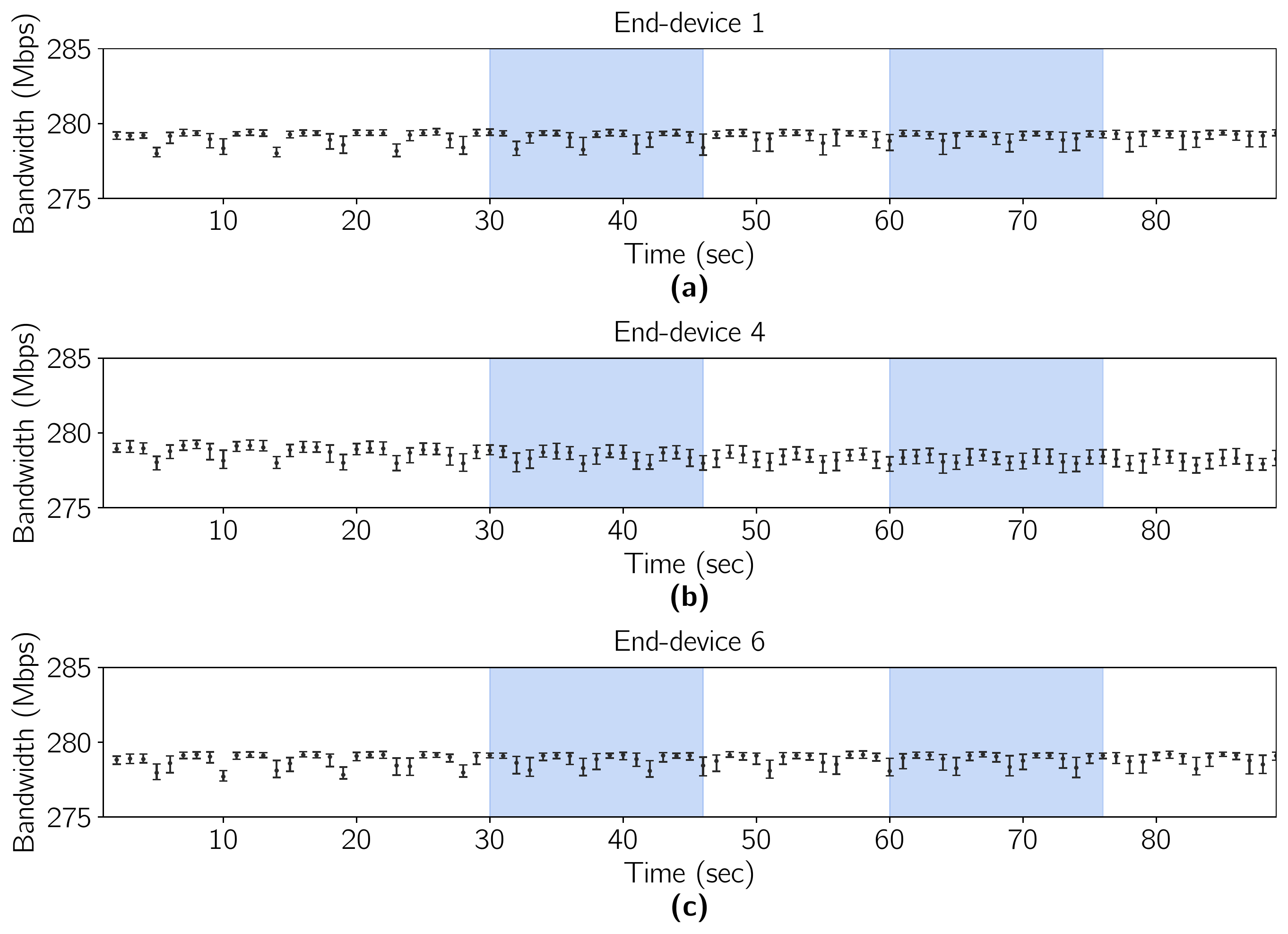}
  \captionof{figure}{Bandwidth readings for end-devices 1, 4, and 6 throughout Test 2a (300 Mbps allocation).
    The bars show the sequential execution of \textit{sleep-app} by 7 end-devices.
  These results show that there is no additional variation in bandwidth for running fog applications in the presence of \textit{sequential} service requests made to the FDK by other end-devices. 
  }
  \label{test2a}
\end{figure}

\textbf{Test 2a.} 
Here, the \textit{sleep-app} runs occur \textit{sequentially} with a 2-second gap in between each run.
Figure \ref{test2a} shows the median value, as well as the upper and lower quartile values, for all 90 bandwidth readings of end-devices 1, 4, and 6.
%
%
Although additional messages are flowing through the network at around 30-45 seconds and 60-75 seconds, the transmission speed of \textit{iperf-app} is not affected, indicating that the bandwidth allocations are not compromised by the overhead incurred by the other \textit{sleep-app} runs performed during this time.

\begin{figure}[t]
\centering
\includegraphics[width=0.84\linewidth]{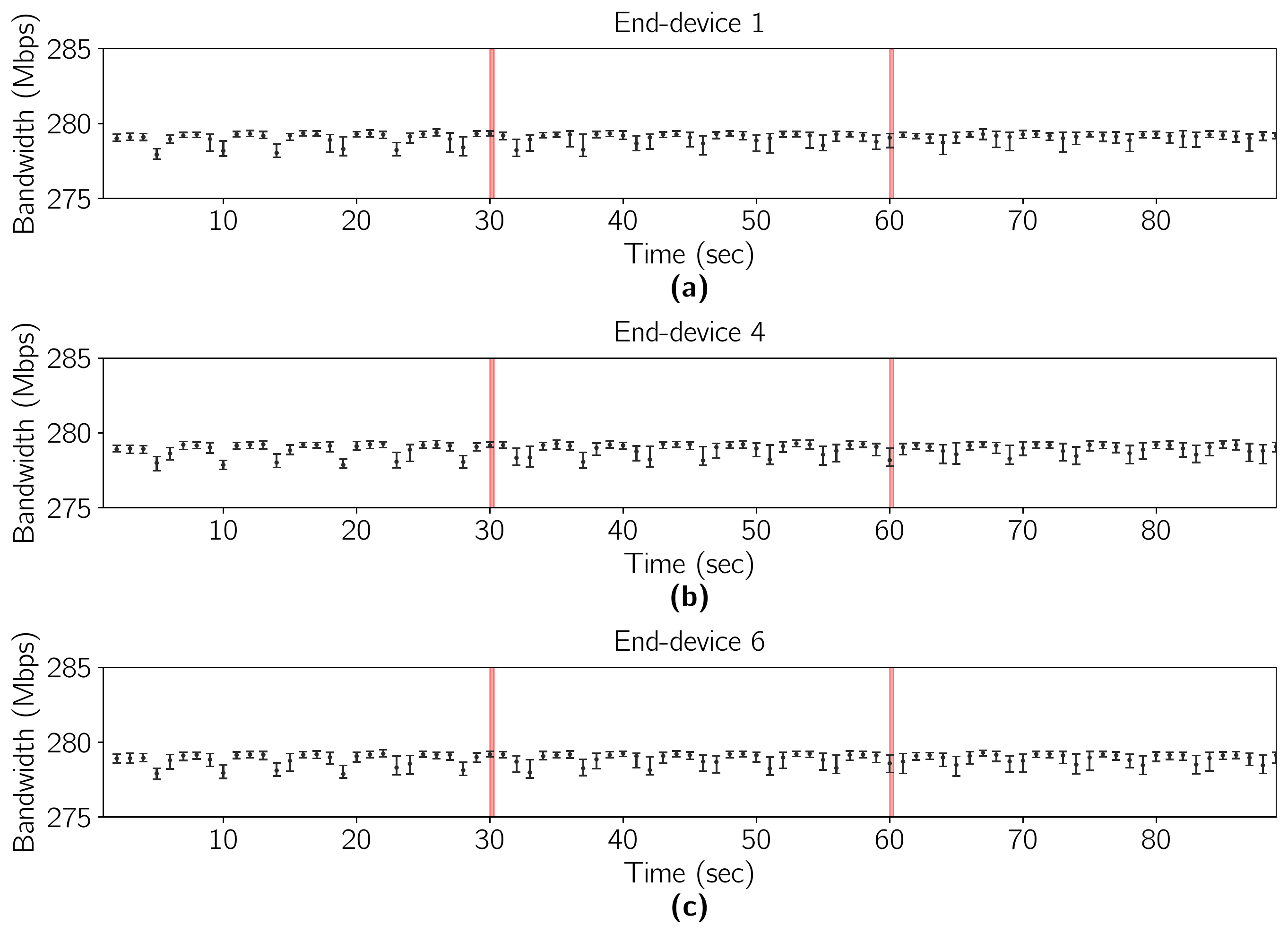}
  \captionof{figure}{Bandwidth readings for end-devices 1, 4, and 6 during Test 2b (300 Mbps allocation).
  The vertical lines show the instances the 7 end-devices start \textit{sleep-app} concurrently.
    These results show that there is no additional variation in bandwidth in the presence of \textit{concurrent} service requests made to the FDK by other end-devices. 
  }
  \label{test2b}
\end{figure}

\textbf{Test 2b.} This test is identical to Test 2a, except that the \textit{sleep-app} runs occur after 30 and 60 seconds into transmission are executed \textit{concurrently}.
Figure \ref{test2b} presents the results.
Similar to the results of Test 2a, we see that there is essentially no drop or variation in bandwidth.
%
%



\subsubsection{Test 3: Multiple Bandwidth Guarantees}

In this test, we evaluate the effect of a large amount of concurrent requests on the service and transmission speeds of multiple fog applications running in parallel.
We subject the hardware to a stress test to measure how the FDK operates under large volumes of requests and to see if bandwidth guarantees can be reliably fulfilled in a highly-congested network.

For this test, we use end-devices 1 through 7 to run \textit{iperf-app} concurrently, and a bandwidth reading is collected per second for 90 seconds.
Then, at 30 seconds and 60 seconds into the 90-second transmission, end-device 8 executes 15 concurrent runs of \textit{sleep-app} at the same time.
This process is repeated 100 times, meaning that 700 \textit{iperf-app} runs and 1500 \textit{sleep-app} runs are performed in total.
Finally, three different variations of Test 3 are executed, where different bandwidth allocations of 100 Mbps (Test 3a), 200 Mbps (Test 3b), and 300 Mbps (Test 3c) are reserved for each \textit{iperf-app} instance, bringing the total number of \textit{iperf-app} and \textit{sleep-app} runs to 2100 and 4500, respectively.


Once the tests completed, we calculated the average of each one-second bandwidth reading across the end-devices in the three groups.
For example, in the case of Group 1, we initially had 3 bandwidth data sets consisting of 100 runs each (one for each of end-devices 1, 2, and 3), where each run consists of 90 bandwidth readings. 
We then took the average of each bandwidth reading (per second) across every run to create a single data set of 100 runs.
Similarly, the same idea applies to the devices and data for Groups 2 and 3.
Note that we did not include end-device 8 in Group 3 for this test because it was performing 15 concurrent \textit{sleep-app} runs and would have experienced a degradation in performance if it were to run \textit{iperf-app} as well.
This is due to the limited networking and processing capabilities of the Raspberry Pi.



\begin{figure}[t]
\centering
\includegraphics[width=1\linewidth]{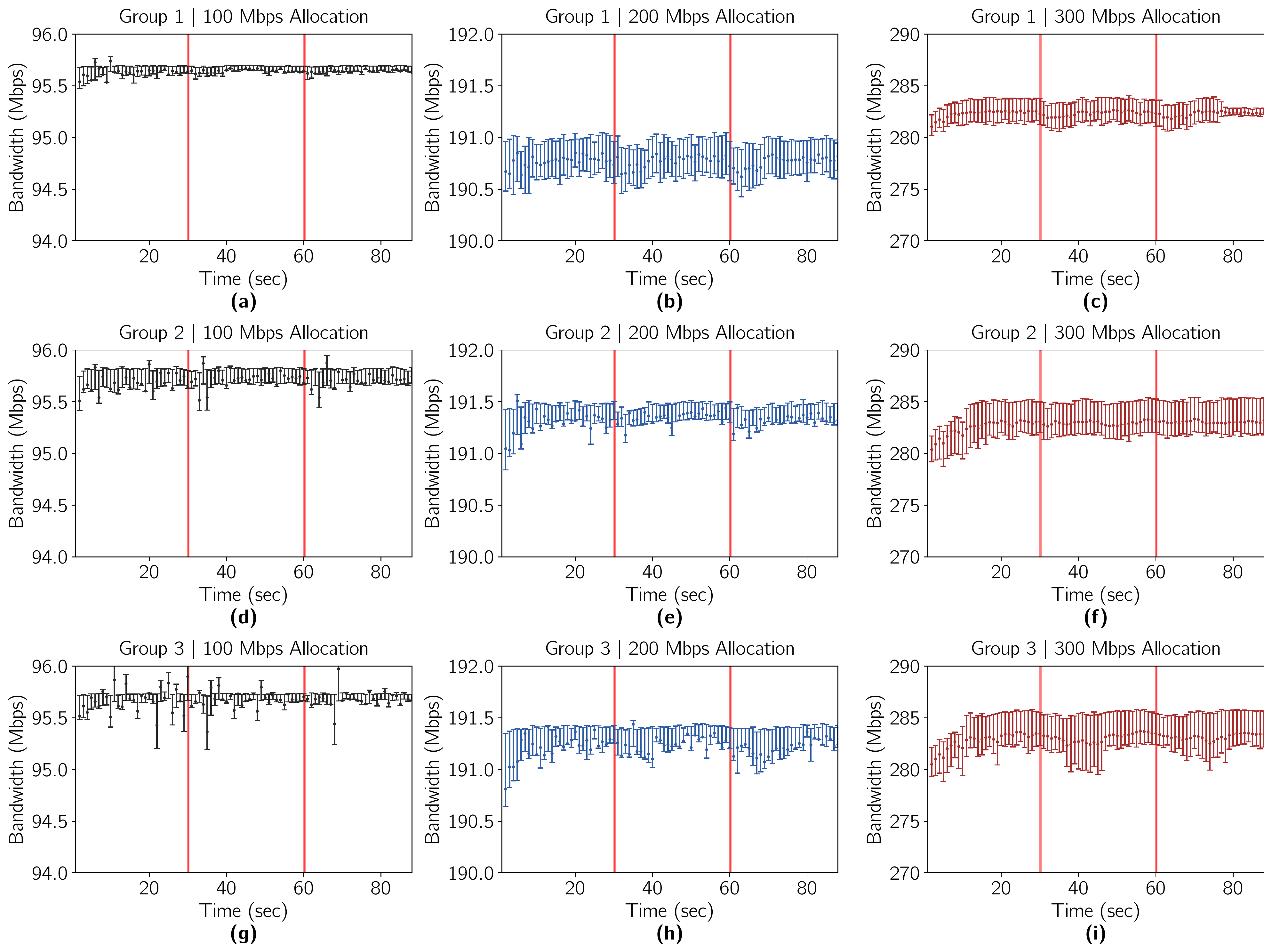}
  \captionof{figure}{
  Actual bandwidth readings for Tests 3a, 3b, and 3c for each Group.
  End-devices 1 through 7 run \textit{iperf-app}, and end-device 8 performs 15 concurrent runs of \textit{sleep-app} at 30 and 60 seconds (as indicated by the vertical lines) into the 90-second \textit{iperf-app} transmissions.
  Even under network congestion and stress during these times, the results show that bandwidth allocations are enforced and no additional variation is observable.}
  \label{test3}
\end{figure}

Figure \ref{test3} shows the results for Test 3.
Here, we formatted the results similar to those of Test 2, where markers for the median value, upper quartile value, and lower quartile value are displayed for each (averaged) bandwidth reading of every run.
Each sub-figure represents all of the data collected for an entire Group.
These results demonstrate less than 1 Mbps variations for 100 Mbps and 200 Mbps allocations, and less than 5 Mbps variations for 300 Mbps allocations.
More specifically, Figure \ref{test3} shows that the actual bandwidth readings are just below the allocated amounts at all times, regardless of traffic stress on the switches.
As previously mentioned, this is because of transmission overheads (such as packet headers) and the limited processing power of the Raspberry Pi boards.



In the case of the 300 Mbps \textit{iperf-app} runs, there are more variations in the bandwidth readings than the 200 Mbps and 100 Mbps runs. 
However, these variations do not correspond to the additional messages flowing throughout the network at 30 and 60 seconds into the 90-second \textit{iperf-app} transmission.
We believe that this is caused by the processing and queuing delays of OVS kernel path. 
Similar observations have been made in \cite{fang2018evaluating}, which confirms that enhancing the switching rate and reducing variations can be achieved by using OVS DPDK and certain GRUB configurations.
We leave these enhancements as future work.

\subsection{Scalability Analysis}
\label{sim_scalability}
A closer look into the operation of the FDK reveals that the five delay components of fulfilling a request are: (i) sending a request from an end-device to the controller, (ii) execution of the RAA to identify a fog-device and a communication path by the controller, (iii) configuration-related communications between the controller and switches and fog-device, (iv) execution of configuration commands on the switches and the fog-device, and (v) sending a reply back to the end-device to confirm the reservation.
Therefore, we can categorize these delays into three groups: \textit{communication delay}: items (i), (iii) and (v), \textit{processing delay of controller}: item (ii), and \textit{processing delay of switches and fog-devices}: item (iv).

The communication delay and processing delay of the controller are affected by network size, which is defined by the number of nodes and the number of links connecting them.
In addition, the communication delay is affected by other factors such as queuing delay and link speed.
The processing delay of configuring switches and fog-devices depends on the hardware and software capabilities of these devices.
In particular, the delay of path reservation on a switch depends on the delays of updating the forwarding table and queue allocation.
Similarly, the delay of fog-device configuration (container instantiation) depends on the processing capabilities of the fog-device.

\begin{figure}[t]
\centering
\includegraphics[width=0.95\linewidth]{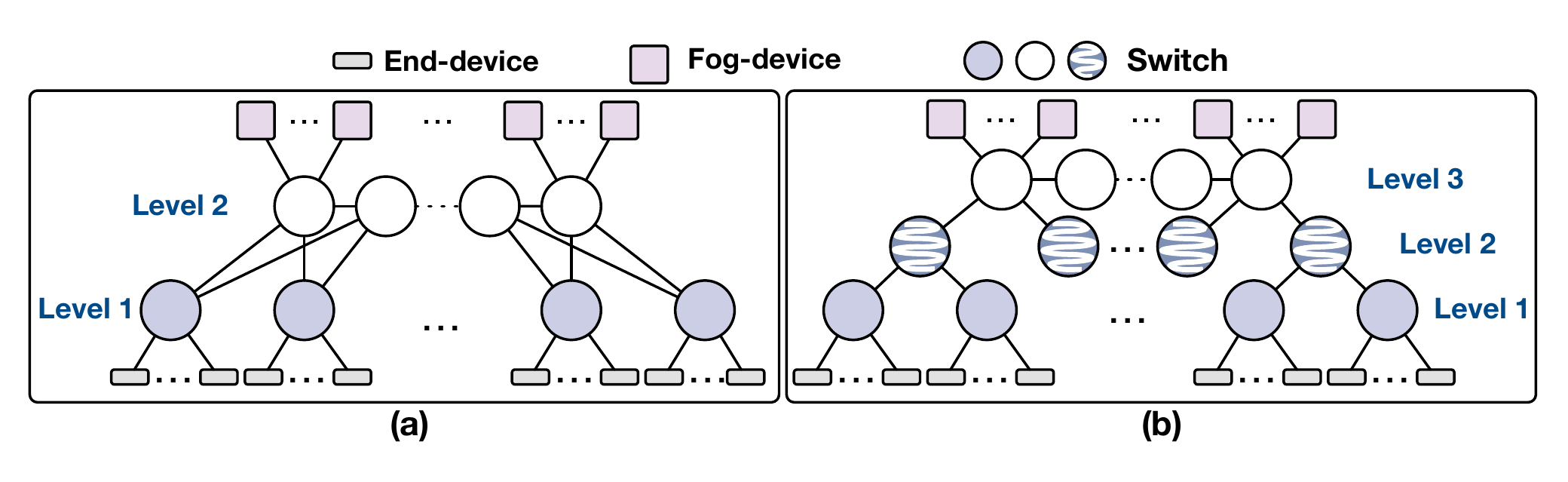}
  \captionof{figure}{The two topologies used for scalability evaluation.
  These two topologies are referred to as 'Topology (a)' and 'Topology (b)' in the text.}
  \label{topos}
\end{figure}

Since fog-device and switch configuration delays depend on the hardware and software characteristics of these components, in this section we neglect these delays and instead focus on the impact of controller processing delay and communication delay on resource allocation.
To evaluate the performance of the FDK versus network size, we developed a simulation tool using the OMNet++ framework \cite{omnet}.
Figures \ref{topos}(a) and (b) present the topologies used, which are inspired by leaf-spine and fat-tree architectures \cite{jyothi2015towards}, respectively.
Topology (a) includes two levels of switches, where each level 1 switch is connected to 1/3 of the (nearest) level 2 switches.
Note that, in order to increase the number of hops between end-devices and fog-devices, we did not connect each level 1 switch to all level 2 switches.
Topology (b) is a tree-like topology that includes three levels of switches, where each level 1 switch is connected to one level 2 switch, and each level 2 switch is connected to one level 3 switch.
The controller is connected to the middle switch in level 2 in Topology (a) and level 3 in Topology (b).
In both topologies, the switches of the highest level are horizontally connected.

\begin{figure}[t]
\centering
\includegraphics[width=1\linewidth]{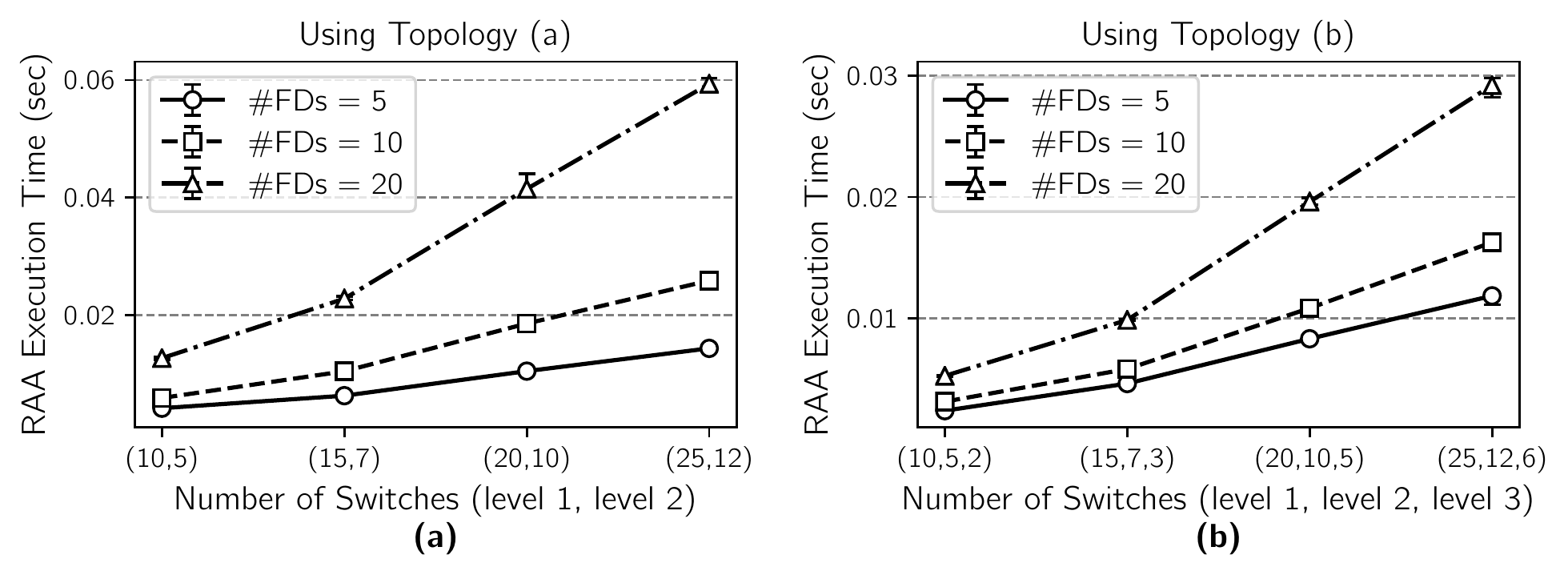}
  \captionof{figure}{Execution time of the RAA (excluding switch and fog-device configuration delay) versus network size and number of fog-devices.
    Sub-figures (a) and (b) present the results for Topology (a) and (b) of Figure \ref{topos}, respectively.
    \#FDs refers to the number of fog-devices per level 2 switch in Topology (a) and level 3 switch in Topology (b).
    The values in each parenthesis on the x-axis refer, from left to right, to the number of level 1, level 2, and level 3 switches.
  }
  \label{RAA_exec_time}
\end{figure}

Figure \ref{RAA_exec_time} shows the RAA execution delay on a single core of a Xeon E5 3 GHz processor.
The time required to evaluate the allocation of resources across all fog-devices to a given end-device is computed, and each point presents the median of these results for all the end-devices.
In other words, referring back to Algorithm \ref{RAA}, we assume that $\mathcal{F}^\prime = \mathcal{F}$, meaning that all fog-devices are eligible to run the service requested by the end-device.
Error bars show higher and lower quartiles.
As discussed in Section \ref{ResourceManager}, the time complexity of the RAA is $O(m\log_{k}n)$.
Also, for a given topology, increasing the number of fog-devices per switch increases the execution time because a higher number of paths must be evaluated whenever a service request arrives.
Increasing the number of fog-devices from 10 to 20 causes approximately a 120\% and 64\% increase in execution time in Figures \ref{RAA_exec_time}(a) and (b), respectively.
Comparing Figures \ref{RAA_exec_time}(a) and (b) shows that the execution time on Topology (b) is about 22\%, 39\%, and 54\% lower than that of Topology (a) when the number of fog-devices per highest level switch are 5, 10, and 20, respectively.
This reduction is because of the fewer number of paths in Topology (b).
For example, although the number of nodes in Topology (b)'s configuration (25,12,6) is higher than that of Topology (a)'s configuration (25,12), the number of paths between each end-device and fog-device is lower in the former topology.

\begin{figure}[t]
\centering
\includegraphics[width=1\linewidth]{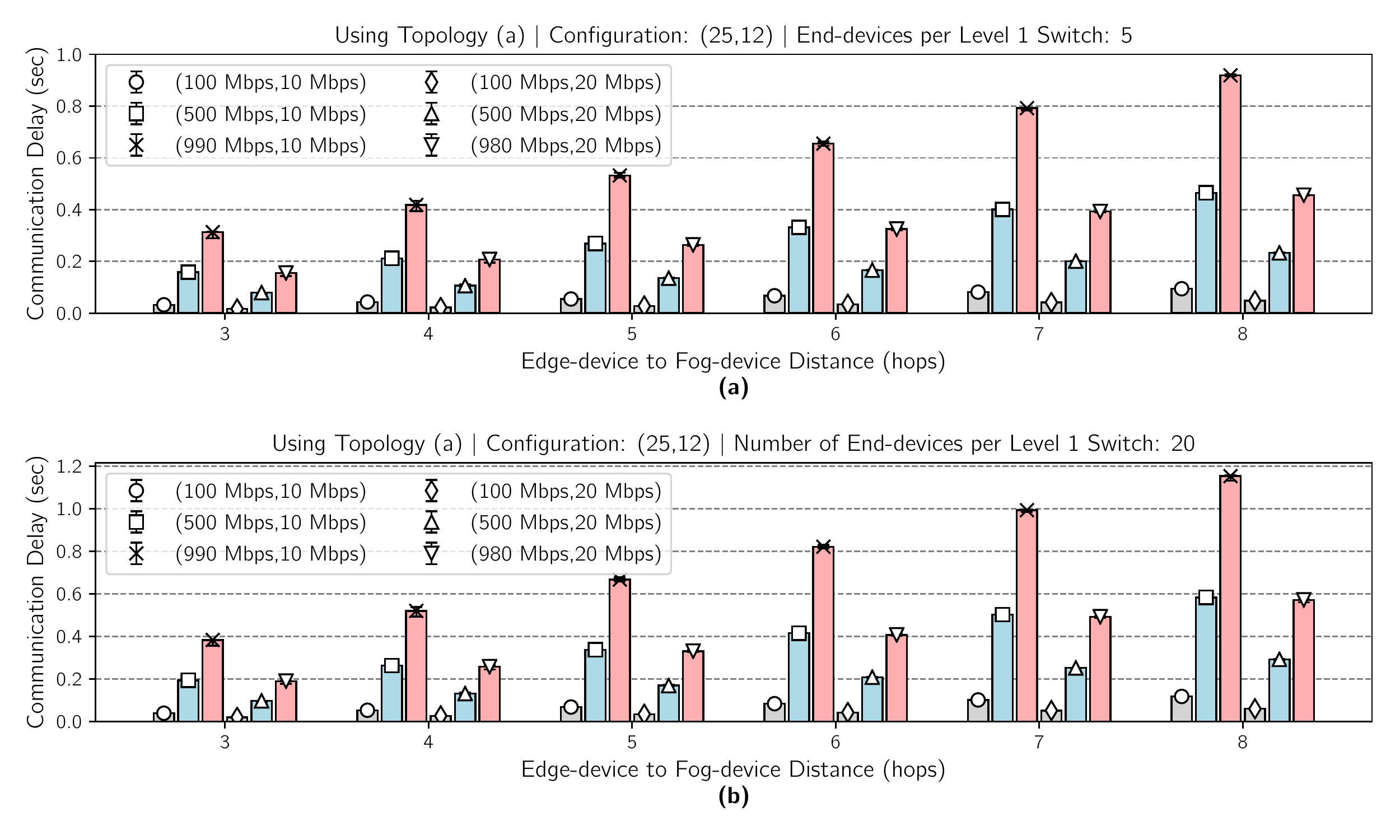}
  \captionof{figure}{Communication delay of resource allocation versus end-device to fog-device distance (hops) for Topology (a) presented in Figure \ref{topos}.}
  \label{alloc_delay_a}
\end{figure}

\begin{figure}[t]
\centering
\includegraphics[width=1\linewidth]{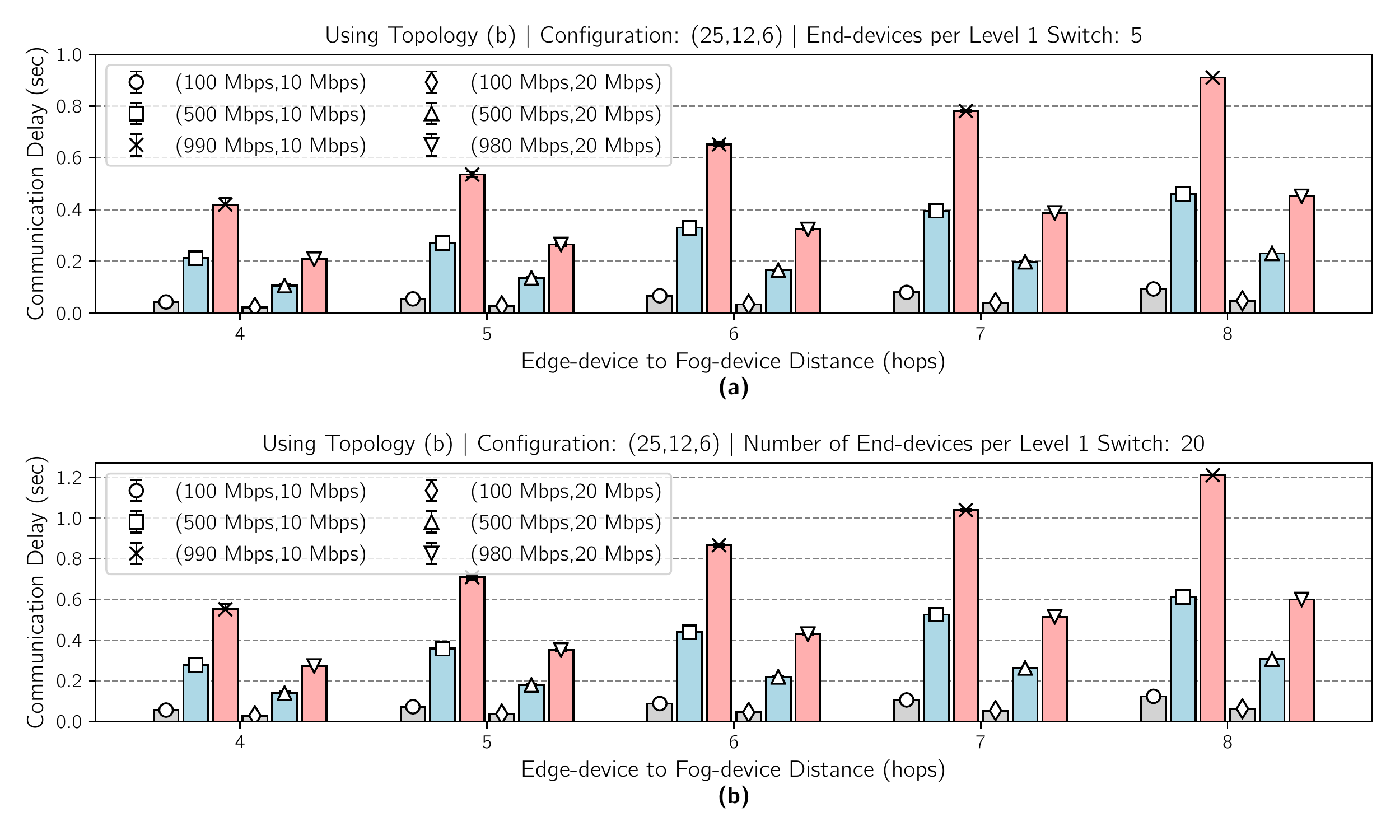}
  \captionof{figure}{Communication delay of resource allocation versus end-device to fog-device distance (hops) for Topology (b) presented in Figure \ref{topos}.}
  \label{alloc_delay_b}
\end{figure}

The next set of results presents the communication delay of resource allocation.
Figures \ref{alloc_delay_a} and \ref{alloc_delay_b} show the median communication delay during resource allocation versus the number of hops between end-devices and fog-devices for all the possible allocations of fog-devices to end-devices.
Each configuration is presented as a tuple ($x$, $y$), where $x$ refers to the bandwidth \textit{used} by all data flows (end-device to/from fog-device) and $y$ refers to the bandwidth \textit{allocated} to the exchange of control flows (items (i), (iii), and (v)) between nodes and the controller.

Both Figures \ref{alloc_delay_a} and \ref{alloc_delay_b} exhibit the impact of the number of hops and background traffic on allocation delay.
The figures show that a higher number of hops increases the number of switches that must be configured along the reservation path.
More specifically, they show that doubling the number of hops doubles the allocation time.
A higher utilization level of links by data flows causes a higher queuing delay on the egress ports of switches.
Queuing delay affects all communication delay components including items (i), (iii) and (v).
For example, a 5x increase in the bandwidth allocated to data flows results in about 380\% higher allocation delay.
In cases of high bandwidth utilization by data flows, these results show that doubling the bandwidth allocated to control flows can cut the allocation delay by half.
However, this introduces a trade-off between resource allocation delay and the communication resources available for data flows.

Figures \ref{alloc_delay_a} and \ref{alloc_delay_b} also reveal that increasing the number of end-devices results in a higher allocation delay.
Specifically, when increasing the number of end-devices from 5 to 20, communication delay is increased by 25\% and 32\% in Topology (a) and (b), respectively.
The cause behind this increase is a higher communication delay (caused by OVSDB) between the controller and switches that grows as the number of flows and queues on each of the switches increases.
For example, the allocation of each queue on a switch inflates the number of bytes exchanged during the resource allocation process as follows: 55 extra bytes are sent from the controller to the switch, and 1000 extra bytes are sent from the switch to the controller.

When discussing Figure \ref{RAA_exec_time} we highlighted that Topology (b) reduces execution time by about 50\%.
In contrast, Figures \ref{alloc_delay_a}(b) and \ref{alloc_delay_b}(b) demonstrate that there is a higher communication delay of path reservation in Topology (b).
This increase is about 5\% when the number of end-devices is 20, and it is further increased for a larger number of end-devices (not shown in the results).
This is because the lower number of communication paths between the controller and switches in Topology (b) causes a higher queuing delay that intensifies delay component (iii).
This is also the reason behind the large increase in communication delay versus the number of end-devices in Topology (b) (when comparing Figures \ref{alloc_delay_a}(b) and \ref{alloc_delay_b}(b)).
In summary, by putting together the results of Figure \ref{RAA_exec_time} and \ref{alloc_delay_b}, when the number of level 1 and level 2 switches are 25 and 12, respectively, Topology (b) results in 30 ms lower RAA execution delay and 22 ms higher communication delay.
It should be noted that, if the number of end-devices surpasses 20, the RAA execution delay would be the same but the communication delay would further increase.

\section{Future Work}
\label{FutureWork}
In this section we present potential future works to extend the FDK.

With regards to network resource allocation, we plan to include transmission delay guarantees by adopting approaches similar to \cite{endtoendDelayGuarantee}.
Furthermore, the FDK does not support resource negotiation with end-devices.
This means that if the amount of resources requested by an end-device exceeds the resources available in the system, the end-device simply receives a failure response message from the FDK and cannot determine which resource demands should be reduced or by how much.
We plan on including available resource information in responses to end-devices to promote flexible and more efficient service requests.
Moreover, it is not immediately apparent how an end-device can calculate what amount of resources is appropriate to request.
For example, determining the actual amount of processing and memory capabilities required to execute a fog application in a timely and efficient manner depends on various factors such as data processing algorithms, data generation rate, and the sensitivity of an application to delays.
As such, a proven, efficient solution to this problem is not immediately apparent and will be the key to enhancing interactions and establishing a greater synergy between end-devices and fog-devices.




In terms of scalability, for large-scale fog-systems with stringent resource allocation deadlines, it is essential to partition the network into regions controlled separately. 
Specifically, we propose to use a local controller in each region.
These local controllers are provided with pre-allocated resources by the main controller, and these resources can be allocated to end-devices immediately.
Each local controller can also request (from the main controller) for more resources based on network dynamics.
In addition, instead of using dedicated boxes, local controllers could be implemented in some switches.
An alternative approach to reducing the execution delay of RAA is to create multiple logical overlay networks based on link delays and bandwidth.
Then, for example, if an end-device is requesting 100 Mbps, only the overlay networks with links satisfying the requested amount of bandwidth will be considered.


As mentioned earlier, we assign a separate rate-limited queue to each egress port along the path identified for a reservation.
For systems including a large number of reservations between end-devices and fog-devices, the use of software switches such as OVS allows the deployment of a higher number of flows and queues in comparison to hardware switches.
In the case of software switches, OVS's mega-flow cache can be employed to aggregate flows.
To this end, instead of flow matching on 5-tuples, multiple flows (sharing a properties such as destination fog-device or egress port) could be aggregated \cite{pfaff2015design}.
However, to efficiently benefit from this feature, RAA (Algorithm \ref{RAA}) must be revised as well.

The FDK opens up vast possibilities for the research and development of fog systems in areas such as image classification, medical monitoring, and industrial monitoring and process control \cite{dezfouli2017rewimo,magid2019image}.
In addition to the enhancement and evaluation of the system's building blocks (e.g., resource allocation algorithms and live container migration), further experimentation can be performed using the FDK to identify the shortcomings of existing solutions as well as developing production-ready solutions.

\section{Conclusion}
\label{Conclusion}

In this paper, we proposed the \textit{Fog Development Kit (FDK)}: A platform for the development and management of fog systems.
The FDK provides a comprehensive resource allocation scheme and stands ahead of other alternatives by enabling both computational and networking resource allocation.
Also, the FDK is application-independent and offers a significantly shorter and simplified development cycle for fog-based applications.
In addition to supporting physical, production-grade environments, the FDK significantly reduces development costs by supporting the use of emulation tools as well.
Therefore, the FDK offers applications portability between physical and emulated environments.
These features make the FDK a valuable tool in prototyping and developing any fog system, as they can be created and tested virtually on personal computers and then be easily ported to a physical topology.
These capabilities differentiate the FDK from existing simulation platforms.
By allowing end-devices to request an arbitrary amount of resources and services from fog-devices, the FDK enables the development of large and complex fog systems at essentially no cost, while at the same time abstracting and eliminating the complexity of resource allocation away from developers.








\ifCLASSOPTIONcaptionsoff
  \newpage
\fi

\Urlmuskip=0mu plus 1mu\relax
\bibliographystyle{IEEEtran}
\bibliography{./References}

\end{document}